\documentclass[twocolumn]{aastex62}

\usepackage{graphicx}	
\usepackage{amsmath}	
\usepackage{amssymb}	
\usepackage{multirow}
\usepackage{gensymb}
\usepackage{booktabs}
\usepackage[flushleft]{threeparttable}

\newcommand{\oii}{[O\,{\footnotesize II}]}

\newcommand{\kms}{km s$^{-1}$}
\newcommand{\myr}{$M_{\odot}$ yr$^{-1}$}
\newcommand{\msun}{$M_{\odot}$}
\newcommand{\mstar}{$M_{\star}$}

\received{2019}
\revised{2019}
\submitjournal{ApJ}

\shorttitle{Quiescent galaxies 1.5 billion years after the Big Bang and their progenitors}
\shortauthors{Valentino et al.}

\begin{document}

\title{Quiescent galaxies 1.5 billion years after the Big Bang and their progenitors}


\correspondingauthor{Francesco Valentino}
\email{francesco.valentino@nbi.ku.dk}

\author[0000-0001-6477-4011]{Francesco Valentino}
\affil{Cosmic Dawn Center (DAWN)}
\affil{Niels Bohr Institute, University of
  Copenhagen, Lyngbyvej 2, DK-2100 Copenhagen \O}

\author[0000-0002-5011-5178]{Masayuki Tanaka}
\affil{National Astronomical Observatory of Japan, 2-21-1 Osawa, Mitaka, Tokyo 181-8588, Japan}
\affil{Department of Astronomical Science, The Graduate University for
  Advanced Studies, SOKENDAI, Mishima, 411-8540 Japan}

\author[0000-0002-2951-7519]{Iary Davidzon}
\affil{Cosmic Dawn Center (DAWN)}
\affil{Niels Bohr Institute, University of
 Copenhagen, Lyngbyvej 2, DK-2100 Copenhagen \O}

\author[0000-0003-3631-7176]{Sune Toft}
\affil{Cosmic Dawn Center (DAWN)}
\affil{Niels Bohr Institute, University of
Copenhagen, Lyngbyvej 2, DK-2100 Copenhagen \O}

\author[0000-0002-4085-9165]{Carlos G\'{o}mez-Guijarro}
\affil{Cosmic Dawn Center (DAWN)}
\affil{Niels Bohr Institute, University of
Copenhagen, Lyngbyvej 2, DK-2100 Copenhagen \O}
\affil{Laboratoire AIM-Paris-Saclay, CEA/DSM-CNRS-Universit\'{e} Paris
Diderot, Irfu/Service d'Astrophysique, CEA Saclay, Orme des Merisiers, F-91191
Gif sur Yvette, France}

\author[0000-0001-5983-6273]{Mikkel Stockmann}
\affil{Cosmic Dawn Center (DAWN)}
\affil{Niels Bohr Institute, University of
Copenhagen, Lyngbyvej 2, DK-2100 Copenhagen \O}

\author[0000-0003-3228-7264]{Masato Onodera}
\affil{Subaru Telescope, National Astronomical Observatory of Japan,
National Institutes of Natural Sciences (NINS), 650 North A'ohoku Place, Hilo, HI 96720, USA}
\affil{Department of Astronomical Science, SOKENDAI (The Graduate University for Advanced Studies), 650 North A'ohoku Place,Hilo, HI, 96720, USA}

\author[0000-0003-2680-005X]{Gabriel Brammer}
\affil{Cosmic Dawn Center (DAWN)}
\affil{Niels Bohr Institute, University of
Copenhagen, Lyngbyvej 2, DK-2100 Copenhagen \O}

\author[0000-0002-8680-248X]{Daniel Ceverino}
\affil{Cosmic Dawn Center (DAWN)}
\affil{Niels Bohr Institute, University of
Copenhagen, Lyngbyvej 2, DK-2100 Copenhagen \O}

\author[0000-0002-9382-9832]{Andreas L. Faisst}
\affil{IPAC, California Institute of Technology, 1200 East California Boulevard, Pasadena, CA 91125, USA}

\author[0000-0002-9656-1800]{Anna Gallazzi}
\affil{INAF - Osservatorio Astrofisico di Arcetri, Largo Enrico Fermi 5, I-50125 Firenze, Italy}

\author[0000-0003-4073-3236]{Christopher C. Hayward}
\affil{Center for Computational Astrophysics, Flatiron Institute, 162 Fifth Ave., New York, NY 10010, USA}

\author[0000-0002-7303-4397]{Olivier Ilbert}
\affil{Aix Marseille Univ, CNRS, CNES, LAM, Laboratoire d'Astrophysique de Marseille, Marseille, France}
 
\author[0000-0002-7598-5292]{Mariko Kubo}
\affil{National Astronomical Observatory of Japan, 2-21-1 Osawa, Mitaka, Tokyo 181-8588, Japan}

\author[0000-0002-4872-2294]{Georgios E. Magdis}
\affil{Cosmic Dawn Center (DAWN)}
\affil{Niels Bohr Institute, University of
Copenhagen, Lyngbyvej 2, DK-2100 Copenhagen \O}
\affil{DTU-Space,
Technical University of Denmark, Elektrovej 327, DK-2800 Kgs.\
Lyngby}
\affil{Institute for Astronomy, Astrophysics, Space Applications and
Remote Sensing, National Observatory of Athens, GR-15236 Athens, Greece}

\author[0000-0001-9058-38921]{Jonatan Selsing}
\affil{Cosmic Dawn Center (DAWN)}
\affil{Niels Bohr Institute, University of
Copenhagen, Lyngbyvej 2, DK-2100 Copenhagen \O}

\author[0000-0003-4442-2750]{Rhythm Shimakawa}
\affil{Subaru Telescope, National Astronomical Observatory of Japan,
National Institutes of Natural Sciences (NINS), 650 North A'ohoku Place, Hilo, HI 96720, USA}

\author[0000-0002-9735-3851]{Martin Sparre}
\affil{Institut f\"ur Physik und Astronomie, Universit\"at Potsdam,
Karl-Liebknecht-Str.\,24/25, 14476 Golm, Germany}
\affil{Leibniz-Institut f\"ur Astrophysik Potsdam (AIP), An der Sternwarte 16, 14482 Potsdam, Germany}

\author[0000-0003-3780-6801]{Charles Steinhardt}
\affil{Cosmic Dawn Center (DAWN)}
\affil{Niels Bohr Institute, University of
Copenhagen, Lyngbyvej 2, DK-2100 Copenhagen \O}

\author[0000-0001-6229-4858]{Kiyoto Yabe}
\affil{Kavli Institute for the Physics and Mathematics of the Universe
  (WPI), UTIAS, The University of Tokyo, Kashiwa, Chiba 277-8583, Japan}

\author[0000-0002-9842-6354]{Johannes Zabl}
\affil{Institut de Recherche en Astrophysique et Planetologie
(IRAP), Universit\'{e} de Toulouse, CNRS, UPS, F-31400 Toulouse,
France}
\affil{Univ Lyon, Univ Lyon1, Ens de Lyon, CNRS, Centre de Recherche
Astrophysique de Lyon UMR5574, F-69230 Saint-Genis-Laval, France}

\begin{abstract}
We report two secure ($z=3.775, 4.012$) and one tentative ($z\approx3.767$) spectroscopic confirmations of 
massive and quiescent galaxies through $K$-band observations with
Keck/MOSFIRE and VLT/X-Shooter. The stellar continuum emission, the
absence of strong nebular emission lines and the lack of significant far-infrared
detections confirm the passive nature of these objects, disfavoring
the alternative solution of low-redshift
dusty star-forming interlopers. 
We derive stellar masses of
$\mathrm{log}(M_{\star}/M_\odot)\sim11$ and ongoing star
formation rates placing these galaxies
$\gtrsim 1-2$~dex below the main sequence at their redshifts. The
adopted parametrization of the star formation history suggests that
these sources experienced a strong ($\langle \rm SFR \rangle \sim
1200-3500$~\myr) and short ($\sim 50$~Myr) burst of star formation,
peaking $\sim 150-500$~Myr before the
time of observation, all properties reminiscent of the characteristics of
sub-millimeter galaxies (SMGs) at $z>4$.
We investigate this connection by comparing the comoving number densities
and the properties of these two populations. We find a fair agreement
only with the deepest sub-mm surveys detecting not only the most
extreme starbursts, but also more normal galaxies. We support these
findings by further exploring the \textit{Illustris}-TNG
cosmological simulation, retrieving populations of both fully quenched
massive galaxies at $z\sim3-4$ and SMGs at $z\sim4-5$, with number densities
and properties in agreement with the observations at $z\sim3$,
but in increasing tension at higher redshift. Nevertheless, as
suggested by the observations, 
not all the progenitors of quiescent galaxies at these redshifts shine
as bright SMGs in their past and, similarly, not all bright
SMGs quench by $z\sim3$, both fractions depending on the threshold assumed to define
the SMGs themselves.
\end{abstract}

\keywords{Galaxies: evolution, elliptical galaxies, stellar content, star formation, high-redshift ---
  Sources: submillimeter: galaxies}



\section{Introduction}
\label{sec:introduction}
Decades of investigations allowed astrophysicists to clearly define a
class of ``quiescent'' galaxies in the local Universe. These systems
are typically characterized by their large stellar masses and sizes,
several billion years
old stellar populations, red colors, little to no active formation
of new stars, very limited amount of cold gas and dust, and an
overdense surrounding environment \citep[][]{renzini_2006}. 
The inferred old ages at $z=0$ suggest that these galaxies were
already in place at high redshift \citep[e.g.,][]{gallazzi_2005,
  thomas_2005} and, indeed, a numerous population of
massive, compact ($\sim1$~kpc) quiescent galaxies that are 
Gyrs old already at $z > 2$ has been firmly established from photometry
\citep[e.g.,][]{daddi_2005, trujillo_2006, toft_2007, cimatti_2008,
  cassata_2013, straatman_2014}, and now securely detected with 
spectrographs \citep[e.g.][M. Stockmann et al., in press]{kriek_2009, vandokkum_2009, vandesande_2013,
 toft_2012, gobat_2012, belli_2014, belli_2017a, belli_2017b}. Their extreme stellar densities
suggest that these compact quiescent systems at $z\sim2$ might be the
remnants of an intense burst of star formation triggered by the rapid collapse of
a large amount of gas occurred at $z>4$. In
this scenario, dissipative gas-rich mergers, counter-rotating gas
streams, or disk instabilities would ignite star formation
in high-redshift and dusty star-forming galaxies detectable
at sub-millimeter wavelengths, quickly consuming the gas and
leaving compact and passive remnants 
\citep[e.g.][]{cimatti_2008, barro_2013, toft_2014, zolotov_2015, tadaki_2018,
  gomez-guijarro_2018, gomez-guijarro_2019}.
The matching number densities, sizes, masses, and formation timescales of
sub-millimeter galaxies (SMGs) at $z\sim4-4.5$ and quiescent galaxies at
$z\sim2$ may support this picture \citep[][and references therein]{toft_2014}. 
On the other hand, the mechanism physically responsible for the
cessation of star formation in these
massive systems is still matter of discussion, with several
scenarios still competing (see \citealt{man_2018} for a recent discussion).\\

\noindent
This evolutionary scheme has been recently challenged by the spectroscopic confirmation of
quiescent systems with $M_{\star}\sim10^{11}$~\msun\ above $z>3$ and up to $z=3.717$
(\citealt{gobat_2012, glazebrook_2017, simpson_2017, schreiber_2018b,
  schreiber_2018c}, S18b hereafter, \citealt{forrest_2019}, C. D'Eugenio et al. in preparation),
as part of a substantial population of photometrically selected red galaxies
\citep[e.g.,][to mention recent results]{fontana_2009, ilbert_2013, muzzin_2013,
  straatman_2014, mawatari_2016,
  davidzon_2017, deshmukh_2018, merlin_2018, merlin_2019, girelli_2019, guarnieri_2019}. In at least one case, their quiescent nature has been
initially challenged by sub-millimeter observations
\citep{simpson_2017}, but later confirmed with a high spatial resolution
follow-up, necessary to disentangle the emission of these 
galaxies from nearby companions \citep{schreiber_2018b}. Systematic studies of
larger samples of $z>3$ photometric candidates in the sub-millimeter further support their
average quiescence \citep{santini_2019}.
The extreme masses, stellar densities, old ages, low SFRs, and number
densities appear to be hardly reproducible by hydrodynamical simulations and
semi-analytical models at $z>3$ \citep[S18b]{steinhardt_2016, cecchi_2019}. 
Catching quenched and quenching galaxies at the highest possible redshifts, thus, represents a
formidable tool to test our galaxy formation models and simulations
and, ultimately, the cosmology. For this scope,
studying high-redshift galaxies allows us to better estimate their
ages, as they are
limited by the age of the Universe \citep{belli_2018}.\\

\noindent
Here we report the discovery of three massive objects at $z=3.77-4.01$
with suppressed star formation, followed-up with Keck/MOSFIRE and VLT/X-Shooter longslit
spectroscopy. In one case, the high-quality of the data allowed us to
estimate the stellar velocity dispersion in the highest-redshift
target, opening the way to the study of its stellar dynamics and
structure. We explored such properties in a dedicated companion paper
\citep{tanaka_2019}. Based on the
observed properties, we further investigated the expected characteristics of the
progenitors of our sample of massive quiescent galaxies and compared
them with dusty star-forming
objects from surveys at $z>4$ selected based on their
sub-millimeter fluxes, testing the evolutionary connection suggested
for lower redshift systems. Incorporating information about the previously
confirmed quiescent galaxies at $z\sim3.5$, we compared the number
densities of this population and the putative SMG
progenitors. Finally, we explored the content of the recent \textit{Illustris} TNG
cosmological simulation in order to look for rare
quenched systems at high-redshift and study their connection with
their star-forming progenitors.\\
 
\noindent
This paper is structured as follows. In Section \ref{sec:selection} we
introduce the sample that we followed up spectroscopically, with the
observations described in Section \ref{sec:observations}. We present
the data analysis in Sections \ref{sec:redshift_estimate} to
\ref{sec:quiescence}. In Section \ref{sec:progenitors} we explore the
connection between our sample of quiescent galaxies and their progenitors
at higher redshift, including the view offered by cosmological
simulations (Section \ref{sec:simulations}). Concluding remarks are
collected in Section \ref{sec:conclusions}. 
We assumed a $\Lambda$CDM cosmology with
$\Omega_{\rm m} = 0.3$, $\Omega_\Lambda= 0.7$, and $H_0 =
70$~\kms~Mpc$^{-1}$ and a Chabrier initial mass function
\citep[IMF,][]{chabrier_2003}. All magnitudes are
  expressed in the AB system. 

\section{Selection}
\label{sec:selection}
We based our search for suitable quiescent galaxy candidates on a
combination of full modeling of the optical and near-infrared light
and a rest-frame color-color selection. The former has proved to be a trustable way to
select quiescent objects up to $z\sim2$ (M. Stockmann et al.,
in press) and it naturally incorporates the whole information
available from the photometry, while relying on a set of assumptions
on models and templates. Color-color diagrams are flexible
instruments to broadly separate galaxy populations capturing the main
features with limited observations, but valuable information from the
rest of the spectrum might be discarded from the analysis
(\citealt{merlin_2018} for a recent detailed analysis).

Two members of our team fitted the optical/near-infrared spectral
energy distributions of potential targets: MT modeled galaxies in the Subaru-XMM Newton Deep Field
\citep[SXDS;][]{furusawa_2008} and ID
in the COSMOS field \citep{scoville_2007}. Having been designed for 
independent spectroscopic runs at different facilities and epochs, the
original selection of candidate quiescent galaxies in the two fields
was comparable, but not identical. 

\begin{figure*}
\includegraphics[width=0.48\textwidth]{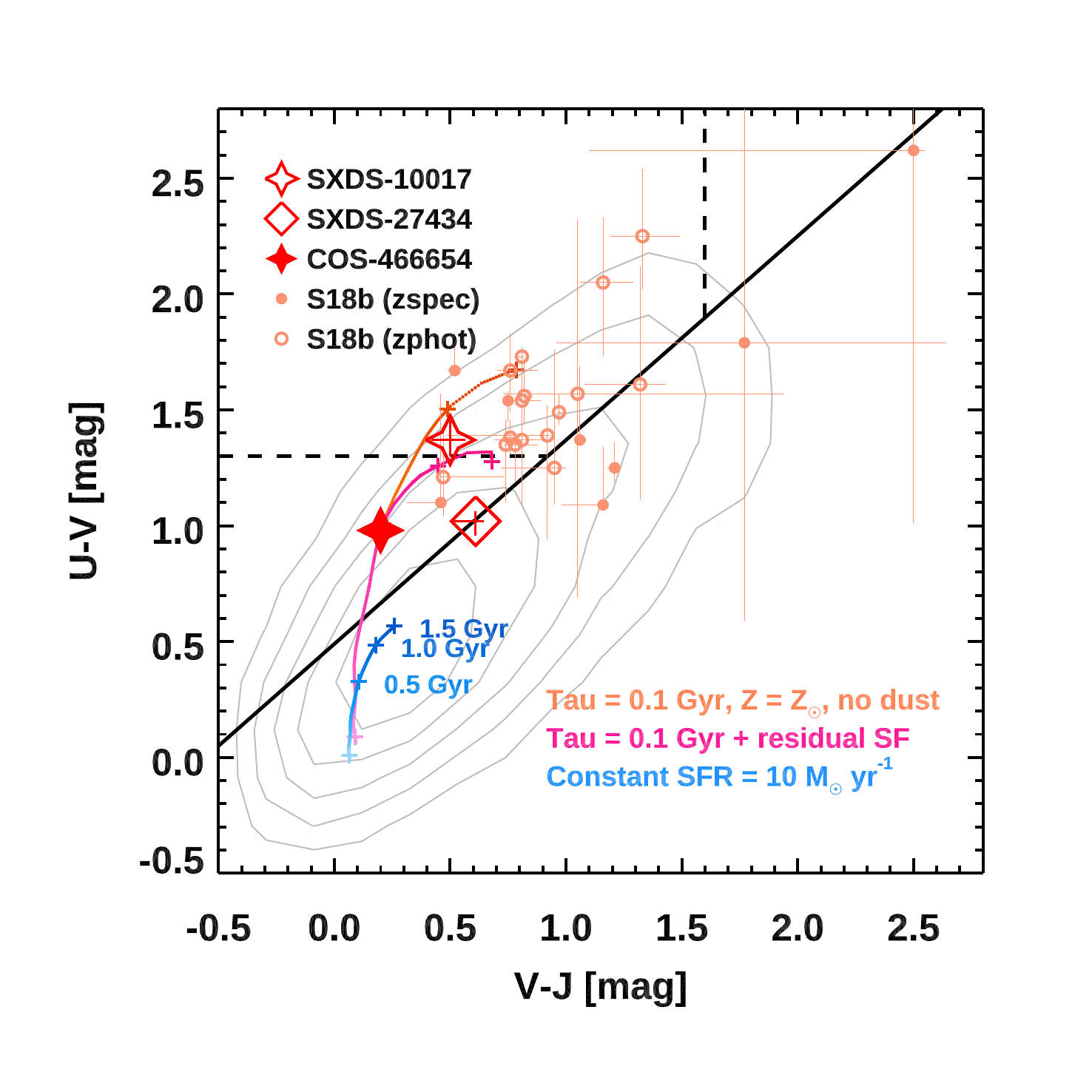}
\includegraphics[width=0.48\textwidth]{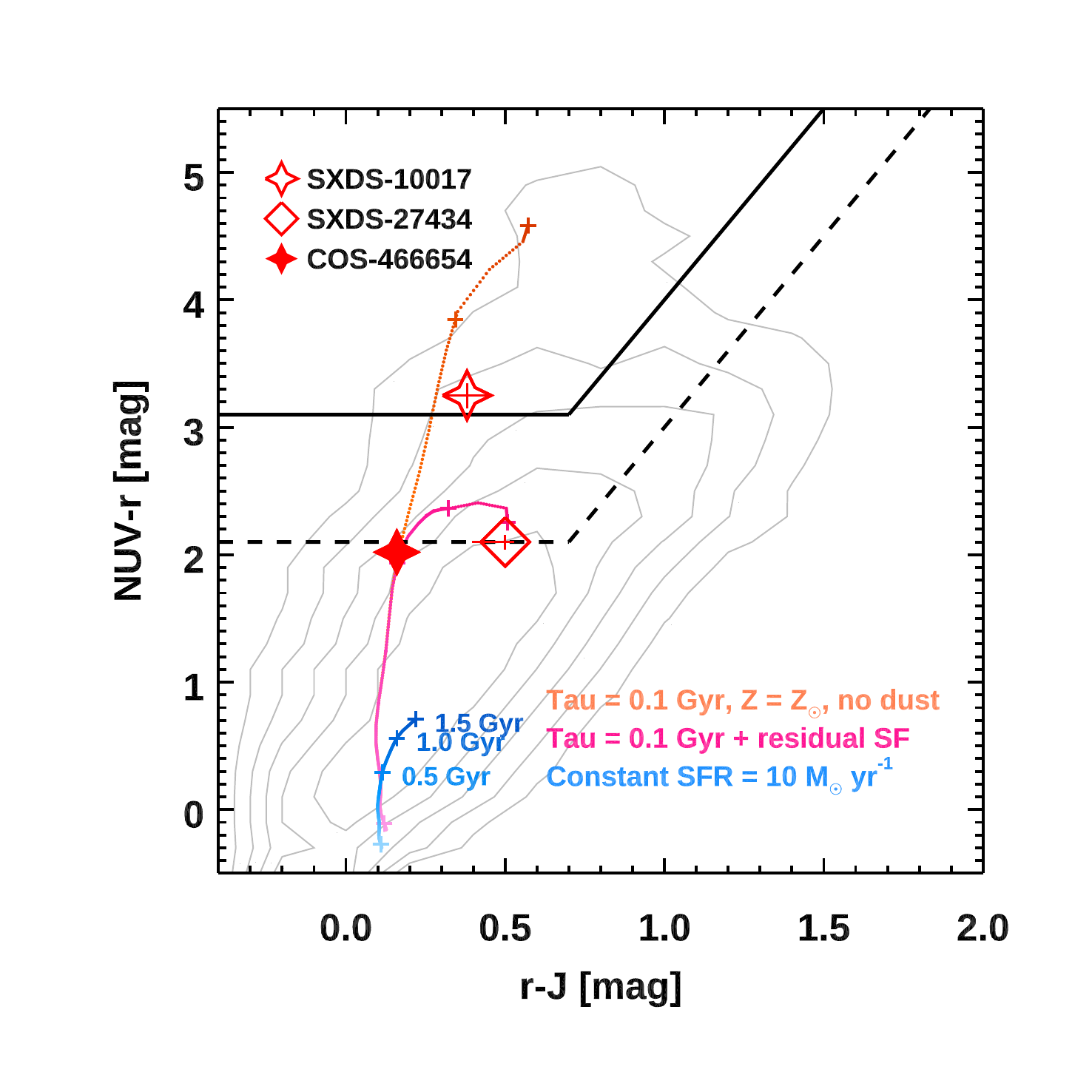}
\caption{\textbf{Rest-frame colors.} \textit{Left:} \textit{UVJ}
  colors for a sample of $3<z<4$ galaxies in the UDS ($7,277$ objects)
  and COSMOS ($11,653$) fields (background gray contours), our observed
  $z=3.77-4.01$ quiescent objects
  (red filled and open stars, red open diamond), and the spectroscopic and photometric
  sample from \citet[filled and open orange circles,
  respectively]{schreiber_2018c}. The background
    contours show the density of UDS+COSMOS points from the parent samples 
  in bin of 0.2 mag on both axes, smoothed over $2\times2$ bins with a
  boxcar average filter. \textit{Right:} \textit{NUVrJ} colors for the
  same objects in the left panel. The same symbols and colors apply
  here. The background contours trace the density of UDS+COSMOS points from the parent samples
  in bin of 0.2 mag on both axes. The tracks in both panels show the
  evolution of synthetic \citet{bruzual_2003} models of rapidly quenched galaxies
  (exponentially declining SFH with $\tau=0.1$~Gyr, $Z=Z_\odot$, no
  dust, orange line), with possible residuals of late star formation
  ($1$ \myr\ after quenching for a galaxy of $10^{10}$
  \msun, pink line),
  or systems characterized by a constant SFHs (blue line), the latter
  simulating an active SFG. The plus symbols mark the time steps as
  indicated by the blue labels, identical for all the tracks.}
\label{fig:rfcolors}
\end{figure*}

\subsection{SXDS/UDS field}
\label{subsec:uds_sample}
We performed the SED modeling for galaxies in the SXDS field as described in
\cite{kubo_2018}. Briefly, we ran the custom Bayesian photometric code
\textsc{Mizuki} \citep{tanaka_2015} on a multiwavelength catalog
comprising $u$-band observations from CFHT/Megacam, optical $BVRiz$
imaging from Subaru/Suprime-Cam
\citep{furusawa_2008}, $JHK$-bands from the UKIRT Infrared Deep Sky
Survey \citep[UKIDSS;][DR10]{lawrence_2007}, and \textit{Spitzer}
coverage from the UKIDSS Ultra Deep Survey (SpUDS, PI: J. Dunlop),
covering an area of $\sim0.8$ deg$^2$. 
We modeled the SED adopting the \cite{bruzual_2003} models,
exponentially declining star formation histories, solar
metallicities, the \cite{calzetti_1994} dust attenuation law, and the
\cite{chabrier_2003} initial mass function. Emission lines are
included using the intensity ratios from \cite{inoue_2011}.  The Lyman-$\alpha$ escape fraction is assumed to be 0.1. 
Given the constraining power of the data covering the
full SEDs of galaxies, we applied top-hat priors, assigning zero
probability to templates outside observationally motivated
ranges. The parameters include $\tau$
  ($\in[1,100]$ Gyr, for the SFH model), dust attenuation (Calzetti
law, with optical depth in the $V$ band $\in[0,5]$), age
($\in[0.001,14]$ Gyr, excluding ages larger than the age of the Universe at
each redshift), and redshift ($\in[0,6]$), while
the metallicity was fixed to solar.  
We have confirmed that our results
do not significantly change if we apply the physical priors
discussed in \cite{tanaka_2015}.
We then selected a sample of $5$ candidate
``quiescent'' and $12$ ``quenching'' (or
``post-starburst'') galaxies based
on their specific SFRs ($\mathrm{sSFR} \leq 10^{-11}$~yr$^{-1}$ and
$10^{-11} \leq \mathrm{sSFR} \leq 10^{-10}$~yr$^{-1}$),
redshift ($4 < z_{\rm phot} < 4.6$), and reliability of the fit 
(reduced $\chi^2 < 4.5$). The
adopted redshift cut would have allowed us to observe the
$4000$~\AA\ break in the $K$ band, providing a constraint on the age
of the stellar populations. In order to
minimize the exposure time for a spectroscopic
follow-up, we finally selected the quiescent candidate SXDS-10017 at $z_{\rm
  phot}=4.07^{+0.07}_{-0.07}$ and the quenching object
SXDS-27434 at $z_{\rm
  phot}=4.12^{+0.03}_{-0.05}$, the brightest galaxies among our
initial pool of sources ($K=22.5$ and
$21.9$~mag, respectively). Consistently with the sSFR cut and quality
checks, SXDS-10017 falls in the quiescent region of both the
$UVJ$ \citep{williams_2009} and $NUVrJ$ \citep{ilbert_2010} rest-frame
color diagrams (Figure \ref{fig:rfcolors}). On the other hand,
consistently with the looser constraint on the sSFR, SXDS-27434 falls outside the
canonical $UVJ$ and $NUVrJ$ limits for quiescent galaxies
\citep{ilbert_2010, williams_2009},
showing rather blue $U-V$ and $NUV-r$ colors. However, this is not 
unexpected for recently and abruptly quenched galaxies at high
redshift \citep{merlin_2018}. A posteriori, using the spectroscopic or photometric
redshift estimates does not significantly change the location of our
targets in the color-color diagrams. In Figure
\ref{fig:rfcolors} we show the expected
color evolution for galaxies with an exponentially declining 
star formation history ($\tau = 0.1$~Gyr) compared with objects with the same SFH,
but residuals of late star formation (a constant
  $1$ \myr\ after quenching for a galaxy of $10^{10}$ \msun) and active SFGs with a
constant SFH of $10$~\myr. The tracks are based on \citet{bruzual_2003} models
with no dust and a $Z_\odot$ metallicity (see \citealt{belli_2018} for a similar
attempt with different parameters and at lower
redshift). We therefore expected SXDS-27434 to show
younger stellar populations than SXDS-10017, allowing us to probe the 
post-starburst or quenching epoch at $z\sim4$.

\subsection{COSMOS field}
\label{subsec:cosmos_sample}
We re-modeled the SEDs of galaxies in the COSMOS field ($\sim1.8$ deg$^{2}$) with \textsc{LePhare}
\citep{arnouts_1999, ilbert_2006} following
\cite{davidzon_2017}, based on the 30-band photometric catalog by
\cite{laigle_2016}. We adopted \cite{bruzual_2003} stellar population
models, exponentially declining and delayed star formation histories,
solar and subsolar metallicities ($Z=0.4Z_\odot$), and a
\cite{chabrier_2003} initial mass function. For the estimate of the
photometric redshifts, we adopted an SMC extinction law
\citep{prevot_1984} and three different flavors of the
\cite{calzetti_2000} prescription, including the bump at $2700$ \AA.
On the other hand, we used the \cite{calzetti_2000} (allowing for
the bump) and \cite{arnouts_2013} extinction laws when estimating
the physical parameters. We then selected
$2$ quenching candidates with $K_s<23$~mag and $NUVrJ$
colors consistent with a
$\mathrm{sSFR}<10^{-10}$~yr$^{-1}$ (Figure \ref{fig:rfcolors}), excluding solutions at
$\chi^2 > 10 $.
We finally chose the brightest among our candidates at $z>4$
(COS-466654, $z_{\rm{phot}}=4.11^{+0.04}_{-0.1}$ and $K_s=22.26$), so to target the $4000$~\AA\ break in the
$K$ band. In agreement with the 
initial constraint on the $NUVrJ$ colors, COS-466654 falls in the
post-starburst or quenching regions of the rest-frame color diagrams
as in the case of SXDS-27434 (Figure \ref{fig:rfcolors}, Section \ref{subsec:uds_sample}).
\begin{figure*}
  \centering
  \includegraphics[width=\textwidth]{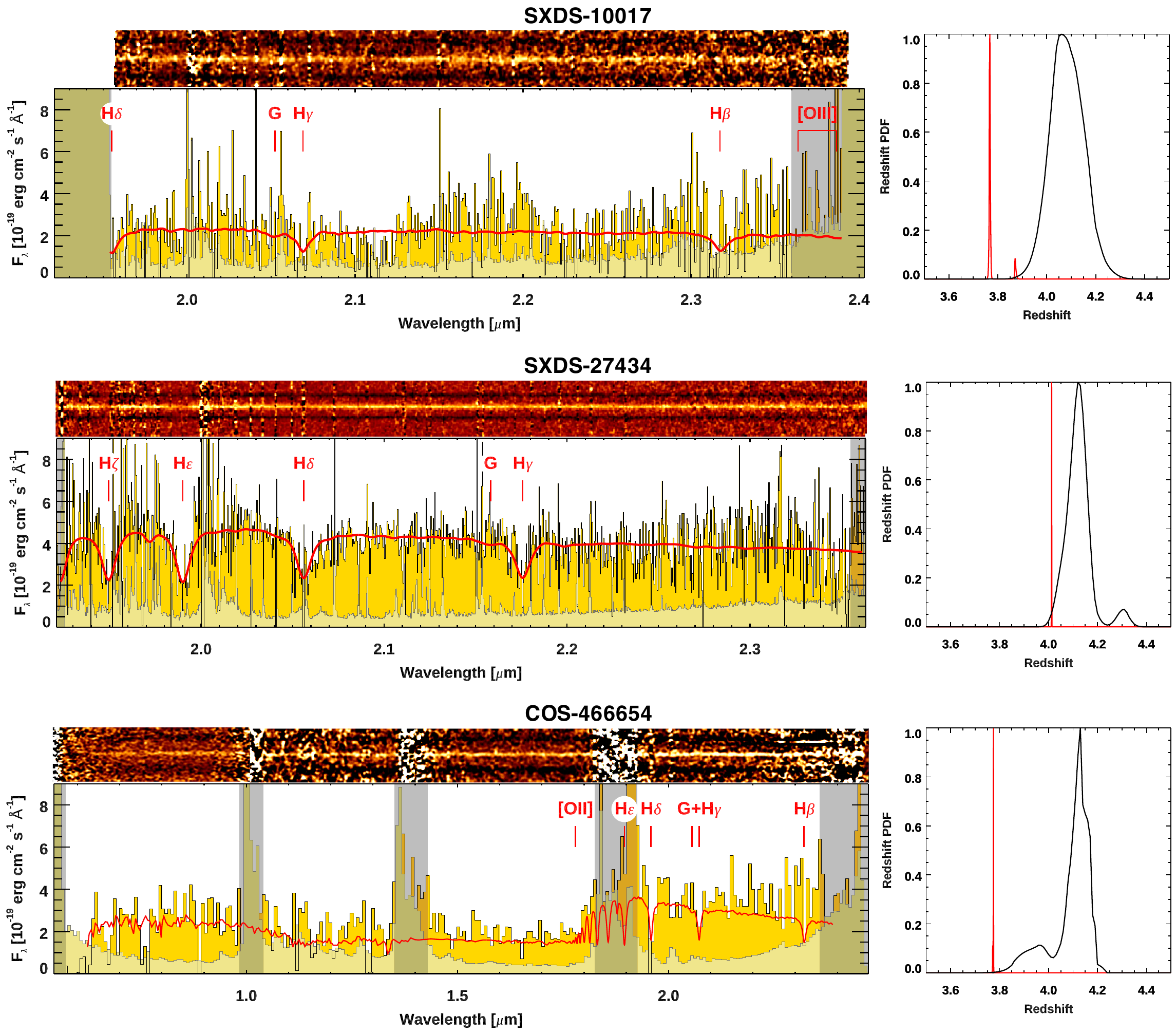}
  \caption{\textbf{Spectra of the quiescent galaxies.} \textit{Top:}
   Keck/MOSFIRE $K$-band spectrum of SXDS-10017, rebinned to a
   wavelength bin of $8.7$~\AA. \textit{Center}:
   Keck/MOSFIRE $K$-band spectrum of SXDS-27434 at its original
   $2.17$~\AA\ resolution. \textit{Bottom:} VLT/X-Shooter VIS+NIR
   spectrum of COS-466654, rebinned to a wavelength bin of
   $72$~\AA. In every panel, the 2D frame is smoothed with a
   Gaussian kernel of 2 pixels of width. The orange
   and gold areas mark the optimally extracted 1D
   spectrum and its noise, respectively. The spectra are rebinned
   differently to reach comparable levels of S/N. The best SED models with
   $z=z_{\rm spec}$ obtained including the $\sim2$ \AA\ resolution
   spectra are shown in red. Each SED model is broadened as mentioned
   in Section \ref{sec:redshift_estimate}.
   The location of the main absorption features
   is labeled. The darker areas indicate the wavelengths of poor
   atmospheric transmission between the observed
   bands. \textit{Right:} The black and red solid lines indicate the probability distribution
   functions for the photometric and spectroscopic redshift, respectively.}
  \label{fig:spectrum}
\end{figure*}

\subsection{Consistency of the applied selection criteria}
As shown in Figure \ref{fig:rfcolors}, the
  rest-frame color and the sSFR selection criteria are
  consistent. In fact, the same  galaxies would have been targeted if
  we applied the COSMOS selection criteria also in SXDS, and vice
  versa. However, we used different SED
  fitting codes to implement the two selections. To test possible
  systematics introduced by this choice, we
  applied the modeling that we used for the SXDS field to the COSMOS
  catalog and vice versa. A 
  small difference in redshifts emerged from this test. For the
  sources in SXDS, \textsc{LePhare} returns
  $z_{\rm phot} = 3.58$ and $4.12$ for SXDS-10017 and SXDS-27434,
  respectively. However, in contrast to the \textsc{Mizuki} selection, the one
  with \textsc{LePhare} does not strictly exclude candidates at
  $z_{\rm phot}<4$, considering the typical accuracy of redshift
  measurements in the distant universe. On the other hand,  
  we retrieve a slightly lower redshift for COS-466654 with
  \textsc{Mizuki} ($z_{\rm phot} =
3.72^{+0.08}_{-0.07}$), but its quiescent state is confirmed. Notice
that the spectroscopic constraints for SXDS-10017 and COS-466654
fall in between the photometric estimates here above (Section
\ref{sec:redshift_estimate}). This kind of
discrepancies may affect statistical studies that purely rely on the parent
photometric samples, but they are less relevant for the present
analysis since we focus on quiescent galaxies individually
confirmed via spectroscopy.

\section{Spectroscopic observations}
\label{sec:observations}

\subsection{Keck/MOSFIRE observations of the UDS field}
\label{sec:mosfire}
We observed the two targets in the UDS field with the Multi-Object
Spectrometer For Infra-Red Exploration \citep[MOSFIRE]{mclean_2012}
at the Keck I telescope on two separate runs. On November
23$^{\rm rd}$ 2017, we collected a total of $4$ hours of observations
of SXDS-10017 out of a complete night initially granted, with
an average seeing of $\mathrm{FWHM}=0.8$'' during the night, as
estimated from individual stars in the field. We
observed SXDS-27434 on December 20$^{\rm th}$-21$^{\rm st}$ 2018 for
7.75 hours, with an average seeing of $\mathrm{FWHM}=0.7$'' over two
half-nights. In both cases, we
observed the targets in $K$-band with a slit width of
$0.7$'', ensuring a nominal initial spectral resolution of $R\sim 3600$. We
adopted the standard ABBA nodding technique with a dithering of
$1.5$'' and 180 s exposures to allow for an optimal background
subtraction. We reduced the data with the MOSFIRE pipeline and
obtained a final flux calibrated, optimally combined 2D spectrum. We
corrected for aperture losses modeling the galaxies as 2D Gaussian curves
with $\mathrm{FWHM}=\mathrm{FWHM_{seeing}}$ and calculating the
light lost outside a 0.7'' wide rectangular slit. The measured
effective major axis of the targets ($R_{\rm eff, maj}=(0.95\pm
0.32)$~kpc and $(0.76\pm 0.20)$~kpc for SXDS-10017 and 27434,
respectively, \citealt{kubo_2018, tanaka_2019}) effectively allows us to treat
the targets as a point sources and to adopt this simple approach. 
We further checked for possible
residual telluric absorptions due to the varying airmass during the
night by calibrating the spectra of several stars observed
simultaneously with our science target, resulting in a negligible
effect. We rebinned the 2D spectra by $2\times, 4\times$, and $30\times$ to a final
resolution of $4.3$, $8.7$, and $65$~\AA, for testing and displaying
purposes. We applied a running optimal weighted mean,
providing the maximal signal-to-noise ratio among the several
approaches we attempted (median, mean, and
clipped mean). We simultaneously increased the noise to account for
the possible
correlation among adjacent spectral elements by forcing the reduced
$\chi_{\rm red}^2=1$ computed in regions of the 2D frame of pure
background. This approach is applicable since the noise variations due
to sky lines occur on scales smaller than the absorption features we aim to detect.
We optimally extracted the spectrum and its
associated noise following \cite{horne_1986}. We finally corrected the
1D spectrum for possible residual flux losses by computing the
synthetic photometry and anchoring it to the best
model reproducing the photometry (Section \ref{sec:sed}). 
The data reduction process for SXDS-10017
resulted in a final median $S/N=6.7$ and a maximum of
$S/N=11.6$ for bins of $65$~\AA, comparable with the performances
reported in S18b for similar integrations and $K$-band luminosities.
Consistently with the brighter $K$-band magnitude
and the $\sim2\times$ longer integration, we find a final median
$S/N=16.4$ and a maximum of $S/N=30.0$ for bins of $65$~\AA\ for
SXDS-27434. The 2D frames and the optimally extracted
spectra are shown in Figure \ref{fig:spectrum}.

\subsection{VLT/X-Shooter observations of the COSMOS field}
\label{sec:xshooter}

We observed the target in the COSMOS field with the cross-dispersed echelle spectrograph
X-Shooter \citep{vernet_2011} mounted on UT2 at the Very Large Telescope
(VLT). The observations were carried out in service mode over
March-April, 2018. A total of 8.6 hours of spectroscopic integration were
spent on target, with an average seeing of
$\mathrm{FWHM}=0.66$''. We observed the target with slit widths of
$1.0$'', $0.9$'', and $0.9$'' in the UVB, VIS, and NIR arms,
respectively, ensuring a lower limit on the resolving power of 4350,
7450, and 5300 in the three arms, given the average seeing smaller
than the slit widths. The chosen configuration and the seeing
conditions allowed us to cover
the wavelength range from $3000$ \AA\ to $24,800$ \AA\ with minimal
slit losses. We adopted the standard ABBA dithering technique with a nod
throw of $4.5$'' and a jitter width of 1''. We optimized the observing
time for 1 hour observing blocks (OB) exposing for $420$, $448$, and
$480$ s in the UVB, VIS, and NIR arms, respectively. For the only OB of
0.5 hours we integrated for $534$, $563$, and $600$ s, in the three
arms. The spectra have been bias-corrected, flat-fielded, wavelength
calibrated, rectified and flux calibrated using observations of
spectrophotometric standards with the VLT/X-Shooter pipeline
\citep{modigliani_2010}. We supplemented the pipeline with optimal frame
combination, telluric correction, and slit flux loss estimate using the
customized scripts described in \cite{selsing_2018} and publicly
available online\footnote{\url{https://github.com/jselsing/XSGRB_reduction_scripts
}}. We rebinned the 2D spectrum using a weighted mean to a final wavelength step of $1.8$ \AA\ in the
NIR arm to match the resolution of MOSFIRE and ensure a
minimum signal-to-noise ratio to measure the redshift. We further
rebinned the spectrum up to $72$ \AA\ for displaying purposes. We
optimally extracted the spectrum \citep{horne_1986} and corrected the
residual flux losses due to seeing variations over the epochs of
observations by computing the synthetic photometry in the available
optical and near-infrared bands and tying them to best model representing the observed
photometry (Section \ref{sec:sed}). The $72$ \AA\ rebinned combined
frame and the optimally extracted spectrum are shown in Figure
\ref{fig:spectrum}. The final median $S/N$ over bins of $72$~\AA\
is $3.3$, $1.7$, $3.8$ and $4.1$ in the VIS arm, \textit{J},
\textit{H}, and \textit{K} bands, with a maxima of $S/N=7.4$, $3.7$,
$6.7$, and $9.4$, respectively. We did not detect significant emission
in the UVB arm.\\

\section{Redshift estimate}
\label{sec:redshift_estimate}
We estimated the redshift using
\textsc{Slinefit}\footnote{\url{https://github.com/cschreib/slinefit}},
a flexible algorithm based on $\chi^2$-minimization that allows for
simultaneous continuum template matching and emission line measurements. We
initially fit the spectra with the best SED models obtained fixing
$z=z_{\rm phot}$. We convolved these models with a Gaussian curve to 
reach a stellar velocity dispersion of $230$ and $214$ \kms\ for
SXDS-10017 and COS-466654 respectively, following the
$\sigma_{\rm vel}$-$M_\star$ relation as in S18b
\citep{belli_2017b}. This is an assumption, as we cannot
constrain the stellar velocity dispersion given the available
combination of signal-to-noise and spectral resolution. On the
contrary, for SXDS-27434 we used the
measured dispersion of $\sigma_{\rm vel}= 268 \pm 59$ \kms\ \citep{tanaka_2019}, consistent with the
$\sigma_{\rm vel}$-$M_\star$ relation we adopted for the other two sources.
We further took into account the possible presence of emission lines with fixed
velocity widths ranging between $60$ and $300$ \kms. However, the
redshift search returns the same results with or without the presence
of emission lines. We explored the redshift range $2<z<5$ and then
refined the redshift measurement within $\pm0.2$ from the best fit
over the large interval. We ran the code on the $\sim2$ \AA\
resolution spectra and on $2\times$ and $3\times$ binned
frames. This did not impact the final solution of
$4.0127^{+0.0005}_{-0.0005}$ for SXDS-27434 consistently with the
one-peak redshift probability distribution (RPD)
and the high probability associated with it ($p=100$\%, integrating the
RPD within $\pm0.01$ from the best fit, \citealt{benitez_2000, brammer_2008};
reduced $\chi^2=1.3$). Using the penalized Pixel Fitting algorithm
\citep[pPXF,][]{cappellari_2004, cappellari_2017} and a slightly different set of
assumptions returns a fully consistent estimate \citep{tanaka_2019}.
Similarly secure is the solution of 
$z=3.775^{+0.002}_{-0.003}$ for COS-466654 ($p=98$\%, reduced
$\chi^2=1.1$). On the other hand, the solution for SXDS-10017
$z=3.767^{+0.103}_{-0.001}$ is more uncertain and it varies by $0.015$
when using the native resolution or the mildly binned spectra. This uncertainty
manifests itself as a second peak of the RPD at $z=3.871$ and with a
lower probability associated with the best solution ($p=81$\%, reduced
$\chi^2=1.6$). We conservatively adopt the solution derived with the
native resolution, despite the
significant improve in probability ($p\sim100$\%) when running
\textsc{Slinefit} on the $3\times$ binned spectrum.
In every case we rescaled the RPD by the empirical factor $C=2$ 
as described in S18b ($P(z) \propto \mathrm{exp} \left[
  (\chi^2 (z) - \chi^2_{\rm min}) / 2C \right]$), in order to take into account
the noise on scales of few spectral elements relevant for the template
matching. We derived independent symmetrical uncertainties randomly
perturbing and refitting the spectrum $1000$ times, obtaining
consistent results ($z=4.0127^{+0.0004}_{-0.0004}$, $z=3.775^{+0.004}_{-0.004}$ and
$z=3.767^{+0.051}_{-0.051}$ for SXDS-27434, COS-466654 and SXDS-10017,
respectively).
We finally refit the spectra with the best SED models
obtained fixing $z=z_{\rm spec}$. In every case we find that our
initial photometric redshift 
overestimated the spectroscopic determination (Figure
\ref{fig:spectrum}). As a result, the initial choice of $z_{\rm phot}>4$
candidates to measure the $4000$~\AA\ break in the $K$ band did not
have success. This is likely due to the choice of following up the
brightest targets, biasing against breaks fully enclosed in
the $K$ band. Based on the criteria defined in S18b, we can 
consider ``robust'' the redshifts for SXDS-27434 and COS-466654, while ``uncertain'' 
the estimate for SXDS-10017. 
\begin{figure*}
  \centering
  \includegraphics[width=\textwidth]{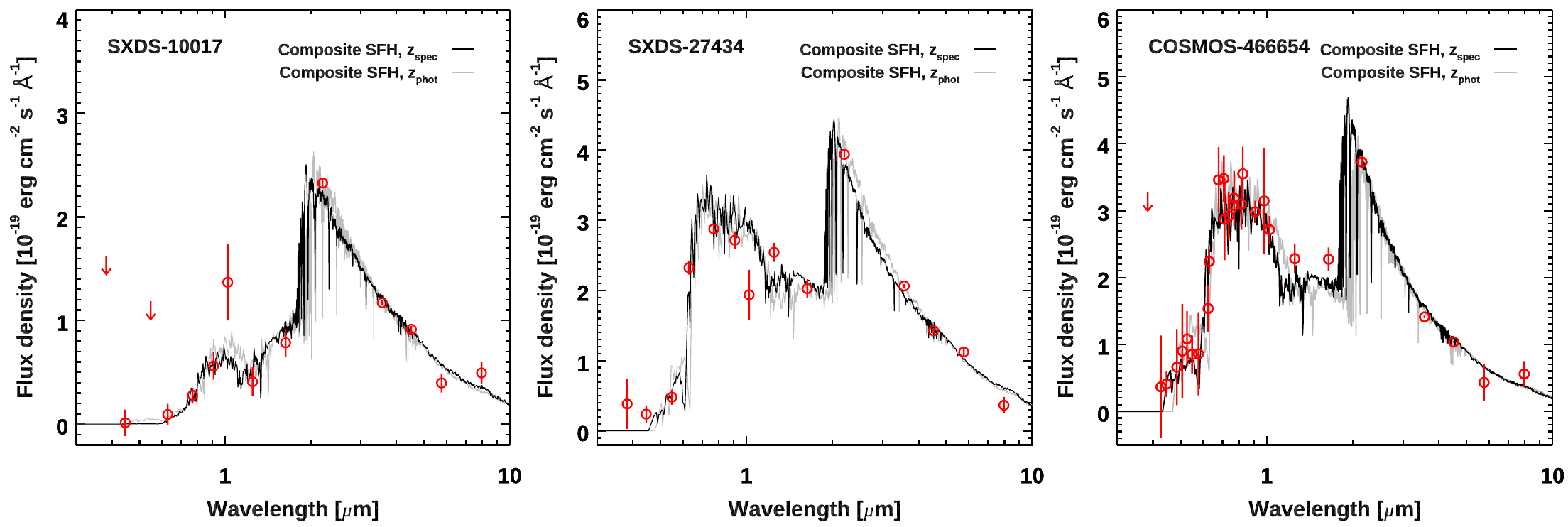}
  \caption{\textbf{Spectral energy distribution of the quiescent
      galaxies.} The gray and black solid lines show the best modeling
    of the spectral energy distribution with fixed $z_{\rm phot}$
    and $z_{\rm spec}$, respectively. In both cases we show the
    results obtained with the composite SFH parametrization by
    \citet{schreiber_2018c}. Red open circles and
    arrows mark the photometric points and $3\sigma$
    upper limits.}
  \label{fig:sed}
\end{figure*}

\begin{figure*}
  \centering
  \includegraphics[width=\textwidth]{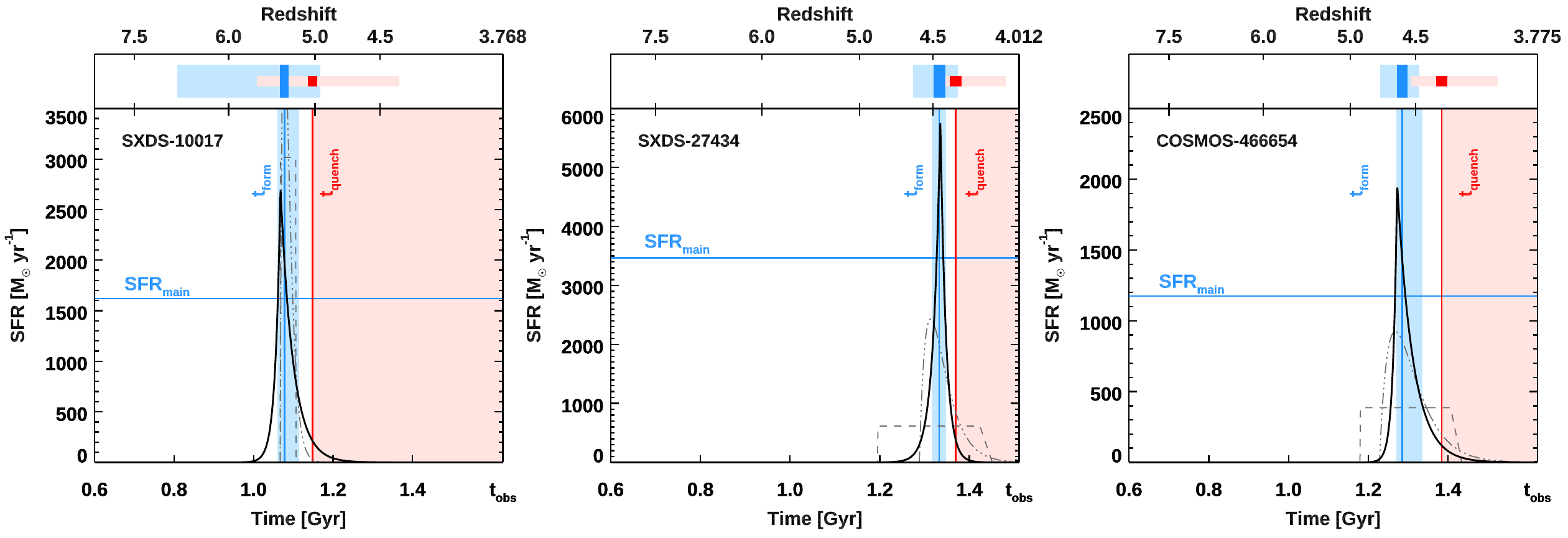}
  \caption{\textbf{Star formation history from modeling of the SED.}
    The black line shows the composite SFH
    (Eq. \ref{eq:composite_sfh}, \ref{eq:composite_sfh2}) corresponding to the best model
    representing the SED in Figure \ref{fig:sed}. The blue vertical line and
    shaded area mark the main formation epoch $t_{\rm form}$ and
    duration of the star formation episode $\Delta t_{\rm form}$ of the
    galaxy. The red line indicates the quenching time $t_{\rm quench}$. The
    blue horizontal line shows the mean $\langle \mathrm{SFR}
    \rangle_{\rm main}$ during the main formation epoch. The dashed
    and dotted-dashed gray lines mark the best truncated and delayed
    SFHs. The blue and red bands in the top insets mark the formation
    and quenching redshifts $z_{\rm form}$ and $z_{\rm quench}$ and
    their 90\% confidence intervals, respectively (Table \ref{tab:properties}).}
  \label{fig:sfh}
\end{figure*}

\section{Modeling of the Spectral Energy Distribution}
\label{sec:sed}
In order to derive detailed physical properties, we re-modeled the
photometry and the rebinned spectra simultaneously with
\textsc{Fast++}\footnote{\url{https://github.com/cschreib/fastpp}},
fixing $z=z_{\rm phot, spec}$ and the maximum possible age to the age
of the Universe $t=t_{\rm
  obs}(z=z_{\rm phot, spec})$. The results are robust against the use
of the photometric or the spectroscopic redshift. We assumed \cite{bruzual_2003} stellar population
models, the \cite{chabrier_2003} initial mass function, and the
\cite{calzetti_2000} dust attenuation law, allowing for extinction
values in the range $A_{\rm V}=0-6$~mag. We
fixed the metallicity to solar $Z=Z_\odot = 0.02$, as reasonable for
very massive objects. We then adopted
multiple analytical parametrizations of the star formation histories
(SFHs):
\begin{itemize}
\item the delayed exponentially declining form $\mathrm{SFR}(t) \propto
  te^{-t/\tau}$ where $t$ is time, widely adopted in
  the literature. We allowed $\tau$ to vary within steps of 0.1 dex
  within $\mathrm{log}\,(\tau / \mathrm{yr}^{-1})= [6.5,10]$ and set a minimum age of $100$ Myr
\item a truncated model, consisting in a constant $\mathrm{SFR} (t)$ over an
  interval $t_{\rm CSF}$ starting at an onset time $t_{\rm onset}$ and then
  instantaneously switched to $\mathrm{SFR}=0$ \myr. The duration $t_{\rm CSF}$ is
  free to vary within $\mathrm{log}\,(t_{\rm CSF}/ \mathrm{yr}^{-1}) =
  [6.5, t_{\rm
  obs}]$ in steps of
  0.1 dex.
\item the composite star formation history (SFH) described in S18b,
so to allow direct comparison with their sample of massive quiescent
galaxies at similar redshifts. This SFH consists of exponentially
rising and declining phases with $e$-folding times free to vary:
\begin{align}
\label{eq:composite_sfh}
\mathrm{SFR}_{\rm base} (t) \propto \left\{\begin{array}{ll}
e^{(t_{\rm burst} - t)/\tau_{\rm rise}} & \text{for $t > t_{\rm burst}$} \\
e^{(t - t_{\rm burst})/\tau_{\rm decl}} & \text{for $t \le t_{\rm burst}$} \\
\end{array}\right.
\end{align}
and $t$ is the lookback time. We adopted the same grid of possible
parameters as in S18b: $t_{\rm burst}=[10\,\mathrm{Myr},t_{\rm
    obs}]$ with (logarithmic) steps of 0.05~dex, $\tau_{\rm
  rise}, \tau_{\rm decl}=[10\,\mathrm{Myr},3\,\mathrm{Gyr}]$ with steps of 0.1~dex.
As in S18b, we further included an extra degree of freedom to decouple
the current SFR from the previous history of formation, allowing for a
burst or abrupt quenching on a short period of duration $t_{\rm free}$
\citep{ciesla_2016}:
\begin{align}
\label{eq:composite_sfh2}
\mathrm{SFR}(t) = \mathrm{SFR}_{\rm base}(t)\times\left\{\begin{array}{ll}
1 & \text{for $t > t_{\rm free}$,} \\
R_{\rm SFR} & \text{for $t \le t_{\rm free}$.} \\
\end{array}\right.
\end{align}
where $t_{\rm free}$ is free to vary between $10$ and $300$~Myr with
steps of 0.5~dex, and $R_{\rm SFR}$ within $10^{-2}$ and $10^{5}$ with
steps of 0.2~dex. 
\end{itemize}

In order to properly compare the results from the different SFH
parametrizations, we computed several integrated quantities
\citep[e.g.,][S18b]{pacifici_2016, belli_2018}. We adopt
the same terminology as in S18b to simplify the comparison between the
two works. We define the epoch of assembly as the ``half-mass formation
time'' $t_{\rm form}$, i.e., the time at which
$50$\% of the total stellar mass was formed, excluding mass loss and
recycling, obtained integrating $\mathrm{SFR}(t)$ over time. The
``duration of the main formation epoch'' $\Delta t_{\rm form}$ is the 
contiguous period enclosing $t_{\rm form}$ and $68$\% of the total
integrated SFR, i.e., limited by the 16\% and 84\% percentiles of the
integral of $\mathrm{SFR}(t)$ over time.
We assumed the mean $\langle \mathrm{SFR} \rangle_{\rm
  main}$ during this period as
representative of ``the typical SFR during the main mass assembly episode''.
We computed ``the quenching epoch'' $t_{\rm quench}$ as the
initial point of the longest contiguous time interval starting from
the time of observation $t_{\rm obs}$ (at $z=z_{\rm spec}$) and
going backwards, where $\mathrm{SFR}<10$\%$\langle \mathrm{SFR}
\rangle_{\rm main}$. We finally adopted the
  uncertainties on the individual parameters estimated by \textsc{FAST++}
  following the $\Delta \chi^2 = (\chi^2
- \mathrm{min}\{\chi^2\})<2.71$ criterion to encompass the 90\% confidence
interval \citep[S18b]{avni_1976, schreiber_2018b}.
We further cross-checked the
uncertainties by bootstrapping $100$ ($1000$) Monte Carlo simulations
for COS-466654 and SXDS-27434 (SXDS-10017) with the same code. The number
of simulations for COS-466654 and SXDS-27434 was limited by the available
computational time, as their spectra have $>2000$ individual
elements. The numerical approach results in less
conservative error bars than the analytical one, as previously found for similar high-redshift quiescent
  objects (S18b). We therefore adopted the
$\chi^2$ criterion as the final estimate for uncertainties
derived from the modeling of the SED.

\begin{deluxetable*}{lccc}
  \tabletypesize{\normalsize}
  \tablecolumns{4}
  \tablecaption{Physical properties of the quiescence galaxies.\label{tab:properties}}
  \smallskip
  \tablehead{
    \colhead{Properties\tablenotemark{a}}&
    \colhead{SXDS-10017}&
    \colhead{SXDS-27434}&
    \colhead{COS-466654}
  }
   \startdata
   Coordinates (RA, Dec)/deg.   & $(34.756250, -5.308038)$      & $(34.29871, -4.98987)$ &
                                                                  $(149.419583, 2.007550)$  \\
      \smallskip
      $z_{\rm phot}$     & $4.07^{+0.12}_{-0.10}$      & $4.12^{+0.09}_{-0.08}$    & $4.11^{+0.06}_{-0.21}$  \\
      \smallskip
      $z_{\rm spec}$     & $3.767^{+0.103}_{-0.001}$      & $4.0127^{+0.0005}_{-0.0005}$ & $3.775^{+0.002}_{-0.003}$ \\
      \smallskip
      $\mathrm{log}(M_\star/M_\odot)$     & $10.89^{+0.05}_{-0.06}$   & $11.06^{+0.04}_{-0.04}$  & $10.82^{+0.03}_{-0.03}$  \\
      \smallskip
      $\mathrm{log(SFR}_{\rm SED}/M_\odot \mathrm{yr}^{-1})$     & $<0.05$    & $1.38^{+0.28}_{-1.25}$ & $0.46^{+0.13}_{-0.69}$  \\
      \smallskip
      $\mathrm{log(SFR}_{\rm H\beta}/M_\odot \mathrm{yr}^{-1})$     & $<0.53$    & ---  & $<0.92$  \\
      \smallskip
      $\mathrm{log(SFR}_{\rm [O\,\scriptscriptstyle{II}]}/M_\odot
      \mathrm{yr}^{-1})$     & ---   & ---  & $<0.41$  \\
      \smallskip
      $A_{\rm V}/$mag     &   $0.2^{+0.2}_{-0.2}$    & $0.7^{+0.1}_{-0.1}$ & $0^{+0}_{-0}$ \\
      \smallskip
      $t_{\rm form} / \mathrm{Gyr}$     & $1.08^{+0.09}_{-0.27}$   & $1.33^{+0.04}_{-0.06}$  & $1.28^{+0.04}_{-0.05}$ \\
      \smallskip
      $z_{\rm form}$     & $5.34^{+1.34}_{-0.37}$    &  $4.51^{+0.16}_{-0.11}$   & $4.64^{+0.15}_{-0.12}$ \\
      \smallskip
      $t_{\rm quench} / \mathrm{Gyr}$     & $1.15^{+0.22}_{-0.14}$    & $1.37^{+0.11}_{-0.05}$  & $1.38^{+0.14}_{-0.08}$  \\
      \smallskip
      $z_{\rm quench}$     & $5.02^{+0.62}_{-0.61}$    & $4.41^{+0.13}_{-0.31}$ & $4.37^{+0.21}_{-0.39}$  \\
      \smallskip
      $\Delta t_{\rm form} /\mathrm{Myr} $     & $50^{+730}_{-27}$   & $32^{+140}_{-10}$   & $56^{+140}_{-41}$ \\
      \smallskip
      $ \mathrm{log(\langle SFR} \rangle_{\rm main} / M_\odot
      \mathrm{yr^{-1}})$           & $3.21^{+0.64}_{-0.40}$      & $3.54^{+0.19}_{-0.35}$ &
                                                                   $3.07^{+1.22}_{-0.30}$ \\
      \smallskip
      $ \mathrm{log}(\tau_{\rm decl}/\mathrm{yr})$& $7.1^{+1.3}_{-0.1}$& $7.2^{+0.2}_{-0.2}$& $7.0^{+0.6}_{-0.0}$\\
  \enddata
  \tablenotetext{a}{The uncertainties on the quantities derived from the SED modeling
    represent the 90\% confidence interval computing following
    \citet[and Section \ref{sec:sed} of this
    work]{avni_1976}. These values are computed adopting
    the composite SFH from \citet[Eq. \ref{eq:composite_sfh} and
    \ref{eq:composite_sfh2} of this work]{schreiber_2018c}.}
\end{deluxetable*}

We show the best-fit SEDs in Figure \ref{fig:sed}, resulting from the
composite SFHs in Figure \ref{fig:sfh}. The best models based on the
three parametrizations of the SFHs are indistinguishable and the
resulting parameters are consistent with each other within the
uncertainties (Appendix \ref{sec:appendix}). Similarly, we tested the choice of the \cite{bruzual_2003}
  models by comparing with the set from \cite{conroy_2010}, retrieving
consistent results within the error bars (Appendix \ref{sec:appendix}).
From here on we therefore adopt the double exponential
SFH as a reference in order to facilitate the comparison with the
sample in \cite{schreiber_2018c}, using the
  \cite{bruzual_2003} templates. The best-fit parameters are reported
in Table \ref{tab:properties}. 
We find that all three sources went through a major burst of star
formation with $ \mathrm{log(\langle SFR} \rangle_{\rm
  main} \, [M_\odot \mathrm{yr^{-1}}]) \sim 3.07-3.54$ ($\langle \mathrm{SFR} \rangle_{\rm
  main}\sim1200-3500$ \myr) over a short period of time $\Delta t_{\rm
  form} \sim 50$~Myr. Notice that this refers only to the last episode of
star formation.
The abrupt quenching following the rapid
formation occurred earlier in time for SXDS-10017 ($z_{\rm quench} =
5.02^{+0.62}_{-0.61}$) than for SXDS-27434 and COS-466654 ($z_{\rm
  quench} = 4.41^{+0.13}_{-0.31}$ and  $4.37^{+0.21}_{-0.39}$,
respectively). This naturally follows our initial selection
and the rest-frame colors, once ascertained the spectroscopic redshift
of the sources and excluded the contamination of low-redshift
interlopers.

\begin{figure}
  \centering
  \includegraphics[width=\columnwidth]{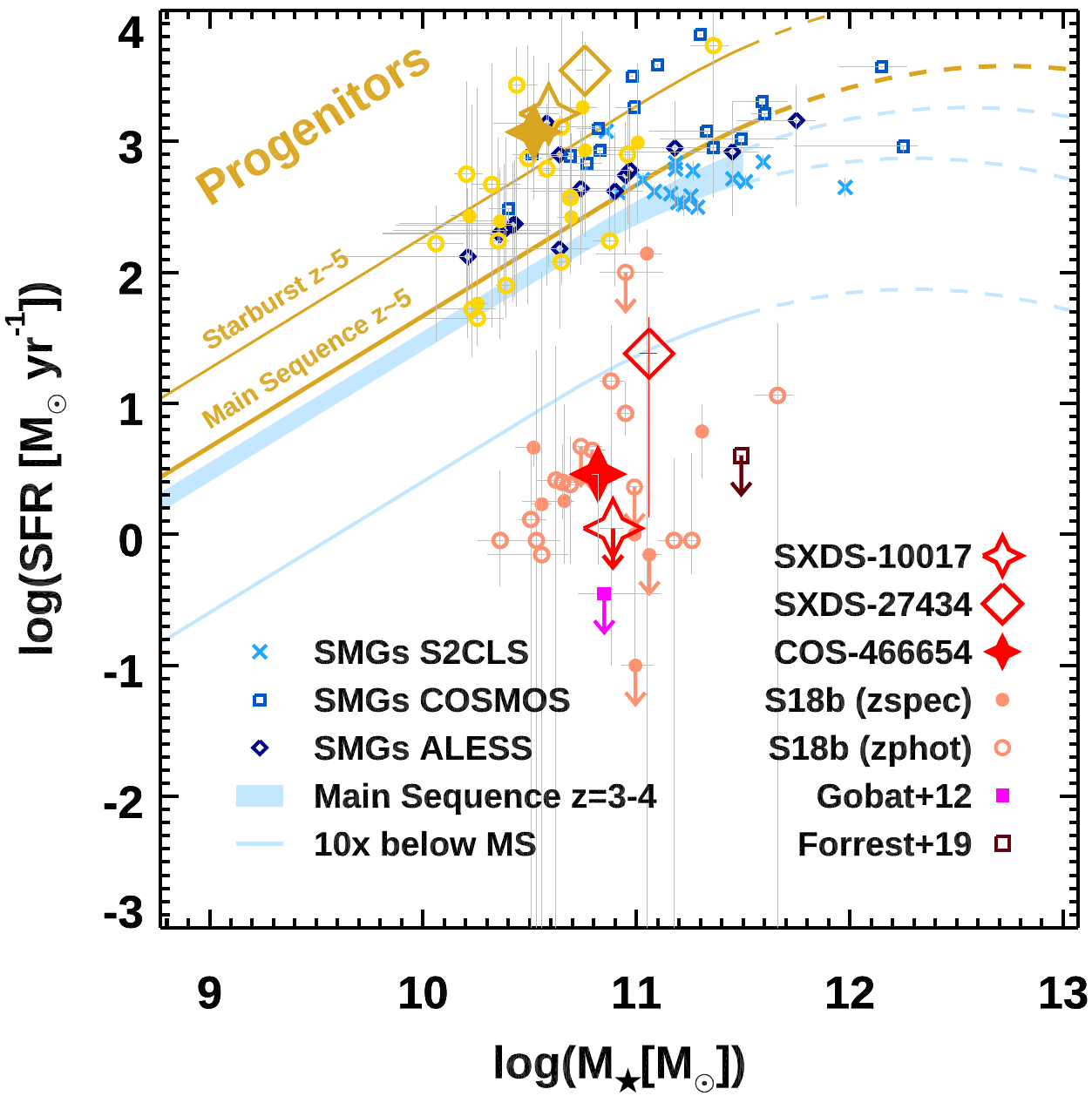}
  \caption{\textbf{Location with respect to the main sequence of galaxies.} The red
  stars and diamond indicate the location of our
  $z=3.77-4.01$ quiescent galaxies in the \mstar-SFR
  plane. The golden symbols mark the $\langle \mathrm{SFR_{\rm main}}
  \rangle$ of their progenitors at $z_{\rm form}$, fixing the mass to 50\% of
  the total final stellar mass.
  The magenta filled square and the brown open square indicate the
  quiescent galaxy at $z=2.99$ and $z=3.493$ reported in
  \citet{gobat_2012} and \citet{forrest_2019}, respectively.
  The filled and open orange circles mark spectroscopically
  confirmed and unconfirmed quiescent sources at $3\lesssim z\lesssim4$ from \citet{schreiber_2018c}. The
  yellow circles show $\langle \mathrm{SFR_{\rm main}}
  \rangle$ of the progenitors at $z_{\rm form}$ from S18b. The location
  of the main sequence at $z=3-4$ as parametrized in \citet{schreiber_2015} is shown by the blue shaded
  area. The thin blue line marks the position of sources $10\times$ below
  the main sequence at $z=3-4$. The golden solid lines indicate the
  position of the main sequence and $4\times$ above it at $z=5\,(
  \sim z_{\rm form})$. The dashed lines show the extrapolation of the main
  sequence to masses larger than $\mathrm{log}(M_\star/M_\odot) = 11.5$. Blue diamonds mark SMGs at $z>4$ from
  \citet{dacunha_2015}, blue crosses from
  \citet{michalowski_2017}, and blue open squares from \citet{miettinen_2017}.}
  \label{fig:ms}
\end{figure}

\section{Quiescence}
\label{sec:quiescence}
From the SED modeling we estimate stellar masses of $\mathrm{log}(M_\star/M_\odot) =
10.89^{+0.05}_{-0.06}$, $10.82^{+0.03}_{-0.03}$, and $11.06^{+0.04}_{-0.04}$ and SFR of 
$\mathrm{log}(\mathrm{SFR}/M_\odot \mathrm{yr}^{-1}) < 0.05$ (90\% upper
limit), $0.46^{+0.13}_{-0.69}$, and $1.38^{+0.28}_{-1.25}$ for
SXDS-10017, COS-466654, and SXDS-27434
respectively. These estimates place the
galaxies $<2.1$, $1.6$, and $1.0$ dex below the main sequence of galaxies at their
redshift, adopting the parametrization of \cite{schreiber_2015}. We
show the location of our targets in the \mstar--SFR plane in Figure
\ref{fig:ms}, along with the sample of similarly selected
massive quiescent galaxies at $3<z<4$ from \cite{schreiber_2018c},
the massive object at $z=3.493$ reported in \cite{forrest_2019},
and the passive galaxy at $z=2.99$ confirmed through \textit{HST}/WFC3 slitless
spectroscopy by \cite{gobat_2012}.

We further estimated an upper limit on the SFR from H$\beta$ and \oii\
emission lines, when covered by our observations. We do not identify
any significant detections from the
line search in the original resolution and the $\sim2$~\AA\ binned
spectra. No evident residual emission appears when subtracting the
best stellar SED continuum model from the spectra.
We therefore put upper limits on the line fluxes as
$\sqrt{\sum_i{\sigma_i}^2}$ where $\sigma_i$ is the noise per $i$-th spectral
bin covered by the potential line. As line
widths we assumed the FWHM of the stellar models we adopted to
estimate the redshift (Section \ref{sec:redshift_estimate}). We then
converted the upper limits into SFR following \cite{kennicutt_1998},
modified according to our \cite{chabrier_2003} IMF. We adopted the
$A_{\rm V}$ extinction from the best SED model to correct for the dust
attenuation. To be more conservative we also computed a final upper
limit using the 90\% upper limit on $A_{\rm V}$ and including a
possible extra-absorption for emission lines following the correction
described in \cite{kashino_2018} ($E_{\rm neb}(B-V) = E_{\rm
  star}(B-V)/0.69$ adopting a \citealt{calzetti_2000} extinction law
for both nebular and stellar emission). For SXDS-10017 we derive
$\mathrm{SFR(H\beta)} < 2.3$ \myr\ ($<3.4$~\myr\ for the $90$\%
conservative upper limit), placing the galaxy $1.8$ ($>1.6$) dex below
the main sequence at its redshift. For COS-466654 we estimate
$\mathrm{SFR(H\beta)} < 8.3$ \myr\ and $\mathrm{SFR([O\,\scriptstyle{II} \textstyle{])}} < 2.6$
\myr\ (90\% confidence interval), corresponding to $>1.2$ and $>1.7$
dex below the main sequence. The $H\beta$ and \oii\
  emission lines are not covered for SXDS-27434, but a similar attempt
  for $H\gamma$ returns an upper limit of $\mathrm{SFR}<8$ \myr\ \citep{tanaka_2019}.

We finally looked for possible far-infrared/sub-mm emission associated
with the three galaxies. SXDS-10017 and SXDS-27434 are not detected in
\textit{Spitzer}/MIPS 24~$\mu$m (SpUDS survey, PI: James Dunlop),
\textit{Herschel}/SPIRE $250$, $350$, and $500$~$\mu$m
bands from the HerMES survey \citep{oliver_2012}, nor in the SCUBA-2
870~$\mu$m maps from the SCUBA-2 Cosmology Legacy Survey
\citep[S2CLS]{geach_2017} or at VLA/1.4 GHz \citep{simpson_2006}. Similarly, COS-466654 is not detected in
any of the mid-infrared ($24$ $\mu$m) to radio (1.4 GHz) bands collected in the
``super-deblended'' catalog of the COSMOS field by \cite{jin_2018},
resulting in a combined infrared signal-to-noise ratio of $\mathrm{SN}_{\rm IR} =
1.8$. At the current sensitivity and spatial resolution limits, this
further confirms the quiescence of the two galaxies and
excludes the presence of bright dusty star-forming companions in their
immediate proximity, at odds with at least one previously reported case
\citep{glazebrook_2017, simpson_2017, schreiber_2018b}.

\section{Progenitors}
\label{sec:progenitors}
After putting on solid ground the existence and the properties of our
targets, we now explore their past history.   
The number of quiescent galaxies at $3<z<4$, their epoch of formation and its
duration, the quenching time, and the average SFR can be used to look
for plausible progenitors. The short formation intervals $\Delta t $ and the
large $\langle \mathrm{SFR} \rangle_{\rm main}$ are reminiscent of the
depletion timescales and the observed SFR of dusty star-forming galaxies (DSFGs)
at high-redshift \citep{daddi_2010,
  tacconi_2010, casey_2014}. To quantify such possible connection, we first
computed the abundance of these two populations in terms of their
\textit{comoving} number
densities. Figure \ref{fig:number_densities} shows a compilation of
values from recent works in the literature. 

\subsection{Number densities of quiescent galaxies at $3<z<4$}
\label{sec:number_densities_qg}
For the quiescent samples, we report the number
densities for $UVJ$-selected galaxies at $M_\star> 4\times 10^{10}$~\msun\ and
$3<z<4$ from UltraVISTA \citep{muzzin_2013} and from ZFOURGE as in
\cite{straatman_2014} (see \citealt{spitler_2014} for an earlier
estimate). The latter has been then updated by S18b after their
spectroscopic follow-up and corrected for contamination of low
redshift interlopers. Notice that S18b computes the number density to a
$25$\% smaller threshold in stellar mass, compatibly with their completeness
limit ($M_\star=3\times10^{10}$ \msun). We then included 
the calculation based on the latest version of the COSMOS
catalog as in \cite{davidzon_2017}, who presented an extensive
comparison with previous works (see references therein); the extended
sample at $3<z<4$ from \cite{girelli_2019};
and
finally from the combined CANDELS fields by
  \cite{merlin_2019}, expanding their previous work on GOODS-South \citep{merlin_2018}. The
authors correct their estimates for incompleteness and
compute the impact of the emission lines on the photometry (see
\citealt{santini_2019} for the recent confirmation of the quiescence
of a subsample of these
galaxies). For consistency, we report the
  estimates with similar mass and redshift cuts as in
  \cite{straatman_2014} and S18b.
We recomputed the number
densities and their uncertainties whenever necessary to match the
criteria above, i.e., by integrating the stellar mass functions in
\cite{muzzin_2013} and \cite{davidzon_2017}. For the other works, we
reported the original values.\\

\begin{figure}
\includegraphics[width=\columnwidth]{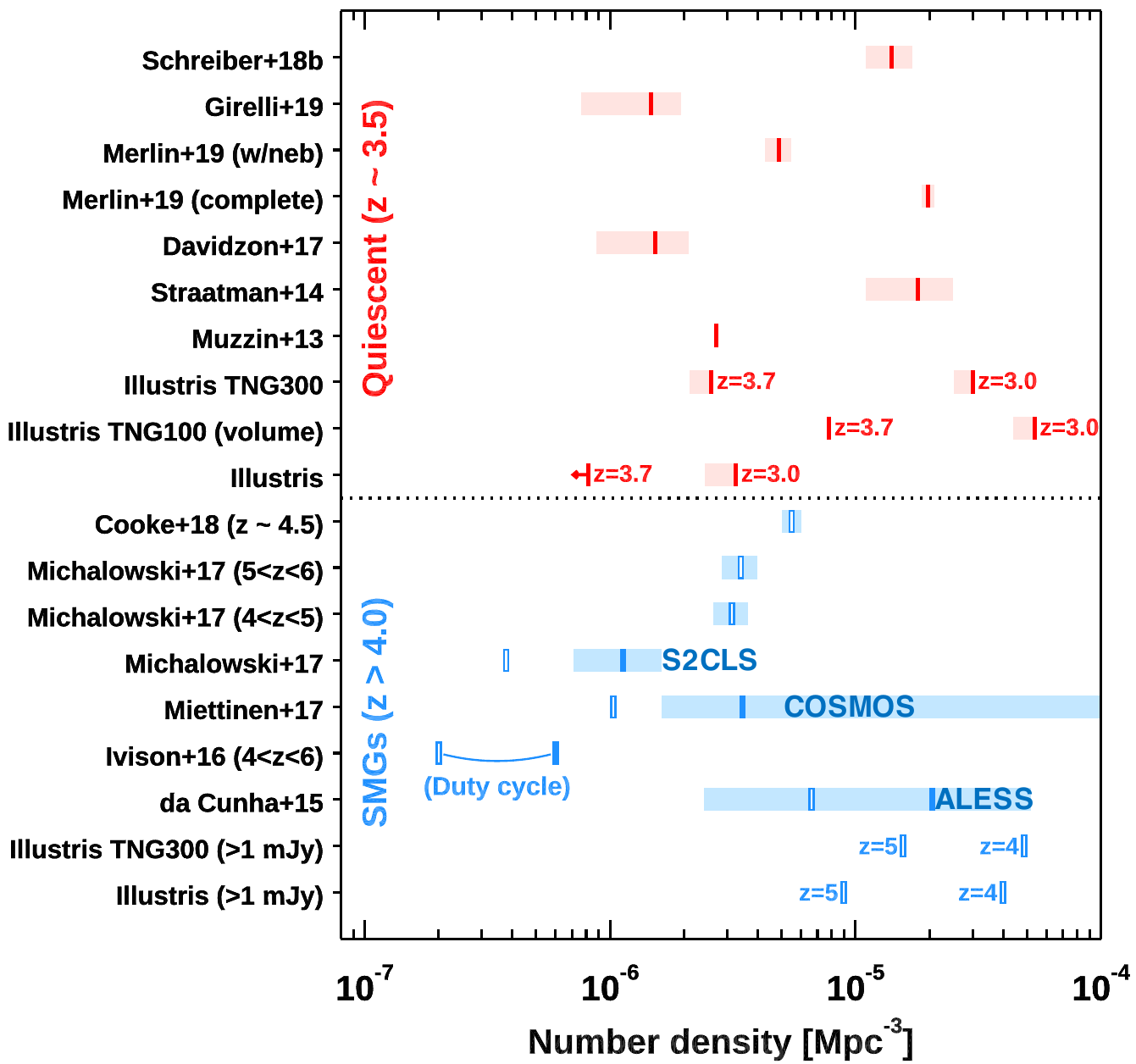}
\caption{\textbf{Number densities of high-redshift quiescent galaxies
    and SMGs.} The red ticks mark the
  observed number densities of massive ($M_\star \gtrsim
  4\times10^{10}$~\msun) quiescent galaxies at $3 \lesssim z \lesssim 4$ from
  the works and projects reported on the Y-axis. The rose areas indicate the
  uncertainties, when available. The blue open ticks mark the number
  densities of variously selected $z>4$ SMGs available in the
  literature. The blue filled ticks include a correction of the duty
  cycle of SMGs ($\rho_{\rm corr} = \rho \times t_{\rm obs}/t_{\rm
    burst}$, where $t_{\rm burst}=200$ Myr), following
  \citet{ivison_2016}. The blue shaded areas indicate the
  uncertainties reported in the original works (Cooke+18,
  Michalowski+17 ($4<z<5$), Michalowski+17 ($5<z<6$)) or recomputed in
  this work based on the $z_{\rm phot}$ uncertainties of the samples
  (Michalowski+17, Miettinen+17, da Cunha+15). The calculations for
  the \textit{Illustris} and \textit{Illustris} TNG (300-1) simulations were
  performed in the snapshots corresponding to the labeled
  redshifts. The comoving number densities of SMGs in both suites are
  based on the catalogs by C. Hayward et al. (in preparation) down to
  a threshold of $1$ mJy, similar to the ALESS limiting flux and
  $\sim3-4\times$ higher than for S2CLS/COSMOS. No duty
  cycle correction is applied to simulations. For the
  simulated quiescent galaxies, the rose area indicates the variation between
  estimates including only ``quenched'' ($\mathrm{sSFR}<10^{-11}$
  yr$^{-1}$) or also ``quenching/post-starburst'' objects ($\mathrm{sSFR}<10^{-10}$
  yr$^{-1}$, red ticks). The values for the
    \textit{Illustris} TNG(100-1) have been corrected by $1.2\times$ for the box
    volume (Section \ref{sec:mass_resolution_effect}).}
\label{fig:number_densities}
\end{figure}

Figure \ref{fig:number_densities} shows a wide range of measurements
for the quiescent population, with variations up to a
factor of $10\times$. The values derived
integrating the stellar mass functions over large area surveys
\citep{muzzin_2013, davidzon_2017} are systematically lower than
computed by counting red galaxies in smaller fields
\citep{straatman_2014, schreiber_2018c, merlin_2019}. We estimated the
impact of the cosmic variance on COSMOS ($1.8$ deg$^2$) and
ZFOURGE-like ($0.1$ deg$^2$) areas as in
\cite{davidzon_2017}, both adopting the analytic approach by \cite{moster_2011}
and by comparing with the mock galaxy catalogs from $24$ realizations in
the Millenium simulations \citep{henriques_2015, springel_2005,
  boylan-kolchin_2009}, the latter including Poisson noise. For masses
$\gtrsim 3\times10^{10}$ \msun, we estimate a $\sigma_{\rm cv}=9.3$\%
($9.8$\%) uncertainty due to cosmic variance from the analytical
(heuristic, including Poissonian noise) approach in the COSMOS field at
$3<z<4$. For the area covered by the ZFOURGE survey for the same
redshift range and \mstar\ threshold, we compute a $\sigma_{\rm cv}=15.3$\%
(a factor of $1.5\times$) uncertainty from the analytical (heuristic,
including Poissonian noise) approach. Besides the variations induced
by cosmic and sample variance, the difference among the various
estimates of number densities is affected by the classification
method, based on colors and/or sSFR with
different thresholds \citep{davidzon_2017, merlin_2019}; the contamination of
lower redshift interlopers (see the discussion in S18b) and AGN
\citep{davidzon_2017}; the slightly different lower
stellar mass integration limits and redshift intervals considered; and
the depth of the observations.

\subsection{A conservative lower limit from spectroscopy}
For the sake of completeness, we finally derived a conservative 
estimate on the comoving number
densities of quiescent objects at $3.219<z_{\rm spec}<4.012$
purely based on spectroscopically confirmed
  galaxies, by combining
our sample and the objects in S18b. Considering the $8$ robust
detections over an area of $2.6$ deg$^{2}$ covered by the COSMOS, UDS,
and ZFOURGE fields\footnote{Two of the ZFOURGE fields are included in
  COSMOS and UDS. Therefore, we did not account for their area in this
calculation.}, we obtain
$n=3.4\times10^{-7}$ Mpc$^{-3}$ at face value and a 3$\sigma$ lower
limit of $>8.8\times10^{-8}$ Mpc$^{-3}$ assuming a Poissonian
distribution (Table 2 in \citealt{gehrels_1986}). Including $5$ extra
uncertain redshift estimates, we derive $n=5.5\times10^{-7}$
and $>2.0\times10^{-7}$ Mpc$^{-3}$ at face value and as a 3$\sigma$
lower limit, respectively. These estimates represent the most
conservative limits on the number of quiescent objects at these
redshifts, being purposely not corrected for any uncertain
completeness effect. With the steady growth of the 
number of confirmed quiescent objects, the constraining power of these
lower limits is destined to rapidly increase in the immediate future.

\subsection{Number densities of sub-millimeter galaxies at $z>4$}
\label{subsubsec:smg_numberdensities}
We collected recent results from large surveys of ``sub-millimeter''
galaxies (SMGs) with detailed modeling of the optical and near-infrared
counterparts, a necessary step to derive at least a photometric
estimate of the redshift and stellar masses. The definition
of a ``sub-millimeter galaxy'' is purely observational and it hides a
certain degree of diversity of the underlying population. However, it
overlaps with the physical definition of DSFGs at the highest redshifts explored so far,
therefore capturing suitable candidates to be the progenitors of massive
quiescent galaxies. In this work we compiled results from recent
surveys on large fields:
\begin{itemize}
\item The (ALMA) Laboca Extended \textit{Chandra} Deep Field South Survey
\citep[(A)LESS,][]{weiss_2009, hodge_2013, simpson_2014,
  dacunha_2015,danielson_2017}: $99$ securely detected sources down to
an rms of $0.4$~mJy
beam$^{-1}$ with ALMA Band 7, originally selected at $870$~$\mu$m
down to rms=1.2 mJy/beam with the
Large Apex BOlometer Camera (LABOCA) on the APEX telescope in the
Extended Chandra Deep Field South field.
The optical/near-infrared follow-up includes 19 bands, with a $3\sigma$ detection limit in
$K_{\rm s}=24.4$ mag \citep{simpson_2014}. Here we use the results of
the full SED modeling with \textsc{Magphys} \citep{dacunha_2008}
presented in \cite{dacunha_2015}. Including the uncertainties on the
$z_{\rm phot}$, $17^{+26}_{-15}$ sources lie at $z_{\rm phot} \geq
4$ over a $0.25$ deg$^2$ area \citep{simpson_2014}. We further take
into account the $2\times$ underdensity of SMGs in the field
\citep{weiss_2009} for the number density calculation. Note that this
area is formally correct only for fluxes covered by
  the original LESS survey. Below this
  limit, the number counts might not be fully representative of the overall
  population of fainter SMGs down to the ALESS detection limit, but
  rather of dim sources in the vicinity of previously known bright
  galaxies \citep{karim_2013}. In this sense, the number density for
  the faintest sources is in principle biased (but see
  \citealt{simpson_2014}). However, in absence of a blind survey with such
  wide coverage, we proceed with the calculation.

\item The SCUBA-2 Cosmology Legacy Survey \citep[S2CLS,][]{geach_2017,
    michalowski_2017}: $\simeq 650$ sources detected at $\geq 4\sigma$
  at $850$~$\mu$m with SCUBA-2 at the James Clerk Maxwell Telescope
  with secure fluxes $\gtrsim 4$~mJy in the COSMOS and UDS fields 
  (area of $2.17$ deg$^2$) and 1.1 mm coverage from ASTE
  AzTEC (see \citealt{dudzeviciute_2019} for a
    recent re-imaging of the UDS field with ALMA).
Roughly $\sim$ 70\% of this sample has a mass estimate
  obtained modeling the optical/near-infrared photometry with
  \textsc{Magphys}, notably assuming a double-component SFH
  \citep{michalowski_2014}. The limiting $3\sigma$ depth in the
  $K_{\rm s}$ band is $24.0/24.9$ mag from the UltraVISTA \citep{mccracken_2012} DR3 deep and
  ultradeep stripes in COSMOS and $25.2$ mag in UDS. Considering only the
  objects above the completeness limit of $\mathrm{SFR}=300$~\myr and taking into
  account the uncertainties on $z_{\rm phot}$, $91^{+63}_{-77}$
  sources lie at $z_{\rm phot} \geq 4$, but only $16^{+7}_{-6}$ with a
  photometric redshift from the optical/near-infrared, the rest being
  determined from the far-infrared SED only.

\item The ALMA follow-up of $124$ SMGs in COSMOS, selected at 1.1 mm
  with ASTE
  AzTEC down to $S^{\rm AzTEC}_{\rm 1.1\,mm}=3.5$~mJy at $>4\sigma$,
  and with optical/near-infrared counterparts
  \citep{brisbin_2017, miettinen_2017}. The ALMA follow-up at 1.3 mm
  reaches an rms of $\sim 0.1$ mJy beam$^{-1}$. The SED has been
  modeled with \textsc{Magphys} to obtain stellar masses and SFR
  \citep{miettinen_2017} and with \textsc{Hyperz}
  \citep{bolzonella_2000} to estimate the photometric redshifts
  \citep{brisbin_2017}. Five sources are spectroscopically confirmed
  above $z>4$, consistently with their photometric redshifts
  \citep{smolcic_2015, gomez-guijarro_2018}. Over a
  covered area of $0.72$ deg$^{2}$, we count $17^{+16}_{-9}$ sources
  with $z_{\rm phot}\geq 4$.  
\end{itemize}

We then computed the number densities of SMGs above
a redshift threshold $z=z_{\rm thresh}=4$ as described in \cite{ivison_2016}:
\begin{equation}
  n = \frac{N}{V_{\rm com}} \mathcal{C}_{\rm duty} \ [\mathrm{Mpc}^{-3}]
\end{equation}
where $N$ is the number of galaxies detected at $z > z_{\rm thresh}$ and
$V_{\rm com}$ is the comoving volume spanned by the observations
detecting the SMG population. $\mathcal{C}_{\rm duty} = t_{\rm
  obs}/ t_{\rm burst}$ corrects for the typically short duty cycle of
SMGs ($t_{\rm burst}$), which shine in the far-infrared/sub-mm regime
only for a fraction of their whole existence or, in this case, of the
cosmic time probed by the observations $t_{\rm  obs}$ \citep[e.g.,][]{toft_2014}. In Figure
\ref{fig:number_densities} we show the number densities 
both uncorrected and introducing $\mathcal{C}_{\rm duty}(t_{\rm burst} =
200\,\mathrm{Myr})$ as a reference (see Section \ref{sec:dutycycle} below).
We included the uncertainties on the
photometric redshift estimates by counting galaxies with $z_{\rm up}$,
$z_{\rm low} > z_{\rm thresh}$, where $z_{\rm up}$, $z_{\rm low}$ are
the upper and lower boundary of the redshift uncertainties as reported
in the original works. This uncertainty dominates the error
budget.  We did not include further corrections for completeness of the
surveys beyond what reported in the literature works (see
\citealt{miettinen_2017} for a caveat for the COSMOS sources). For
reference, we also report two recent results from the literature,
namely number densities for ultra-red galaxies selected from the
\textit{Herschel} Astrophysical Terahertz Large Area Survey
\citep[\textit{H}-ATLAS,][]{ivison_2016} and for a sub-sample of spectroscopically
confirmed SMGs at $z\sim4.5$ from the S2CLS survey \citep{cooke_2018},
along with the number densities originally reported in \cite{michalowski_2017}.\\

Figure \ref{fig:number_densities} shows that large variations are
present in the observed number densities of SMGs. This is likely due to
the variety of selection criteria, the depths, the
spatial resolution, and
the completeness of the various surveys, plus the intrinsic diversity
of the SMG population. The choice of the sub-mm band for the
initial selection corresponds to the sampling of a specific portion of the
far-infrared SED. A selection based on \textit{Herschel} bands or at
$\sim 870$~$\mu$m ($170$~$\mu$m
rest-frame at $z=4$) maps the SED closer to the peak of the dust
emission, being sensitive to the temperature and total IR luminosity
($\propto\,\mathrm{SFR}$). Opting for a cut at $1.1$~mm ($220$~$\mu$m rest-frame at the
same redshift) results into the sampling of the optically thin dust
emission in the Rayleigh-Jeans tail of the SED, thus being mildly
sensitive to the effective temperature and privileging large dust masses over large
SFR, even if ultimately the two quantities are correlated. 
The depth and spatial resolution of
the observations likely has an even stronger impact
than the selection criterion: shallow limits from
  single dish observations allow us to
capture only the rarest and strongest starbursting systems,
likely affected by source blending. Moreover,
the necessity to identify a counterpart in the
optical/near-infrared to estimate stellar masses and redshift 
biases the results against the most dusty and highest redshift
objects. This is evident for the S2CLS sample, for which only
$\sim20$\% of the galaxies with $z_{\rm phot}>4$ have an
optical/near-infrared redshift estimate, the rest being determined from the
far-infrared SED only \citep{michalowski_2017}. However, adopting the
best available photometric redshifts
allows us to improve previous estimates based on the simple assumption
of a fixed fraction of SMGs above $z=4$ \citep[e.g.,][]{straatman_2014,
  ivison_2016}.  
\begin{figure*}
\includegraphics[width=\textwidth]{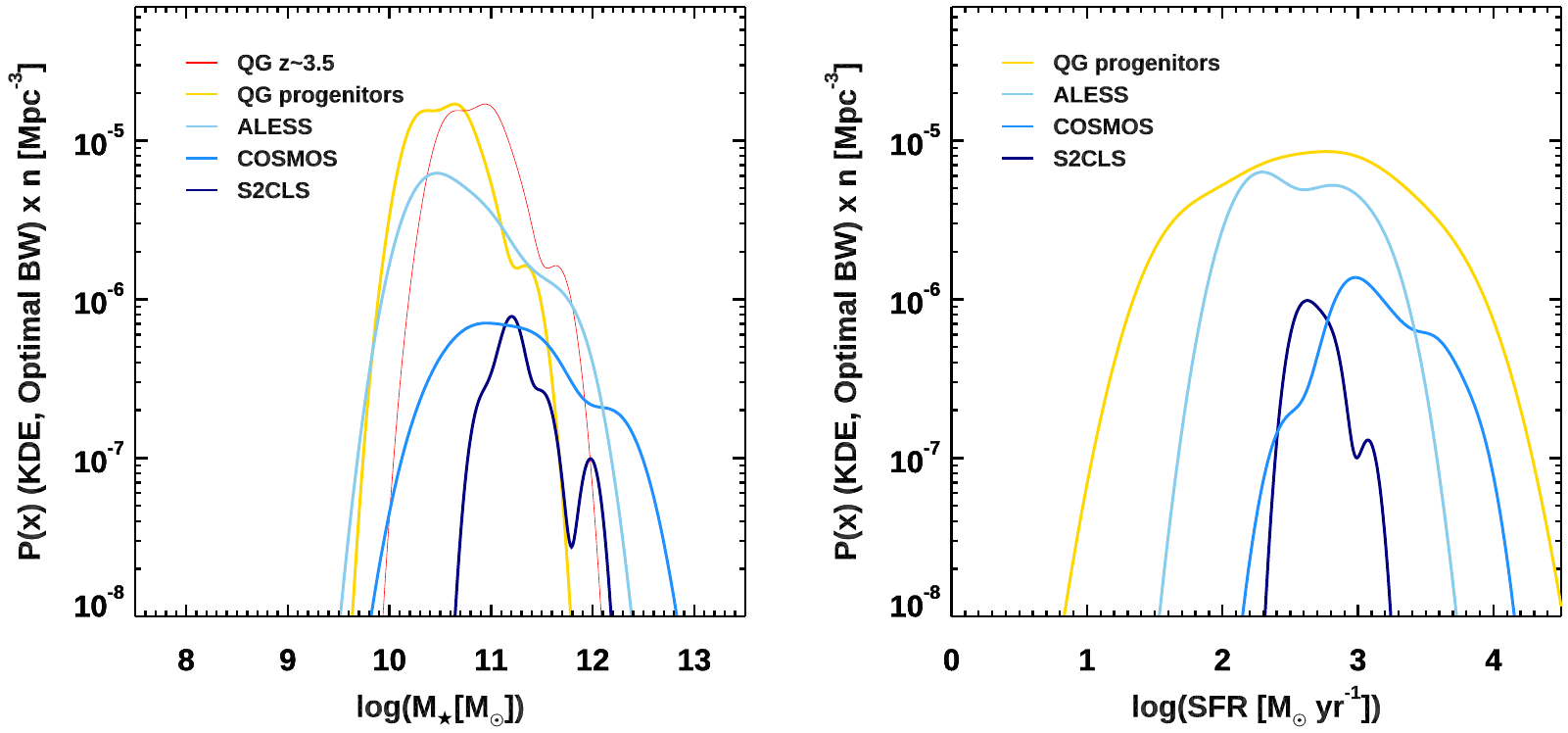}
\caption{\textbf{Stellar masses and SFRs of
    candidate progenitors of high-redshift quiescent galaxies.} Kernel Density
  Estimator (KDE) of the density probability distributions of stellar
  masses (left panel) and SFRs (right panel) for the progenitors at
  $z_{\rm form}$ of massive quiescent galaxies at $3<z<4$ (this
  work+S18b, golden line) and SMGs at
  $z>4$ from the ALESS, COSMOS and S2CLS surveys \citep[blue lines]{dacunha_2015, miettinen_2017,
    michalowski_2017}. For reference, we show the stellar mass
  distribution of the QGs (red thin line). The areas under the curves are normalized to
  the number density of each population (Figure \ref{fig:number_densities}).}
\label{fig:progenitors_properties}
\end{figure*}

\subsection{Standard evolutionary connection}
\label{sec:evolutionary}
We now explore the possible evolutionary connection between SMGs and
quiescent galaxies (see previous results in the literature
at similar or lower redshift \citealt{cimatti_2008, simpson_2014,
toft_2014, straatman_2014, miettinen_2017, gomez-guijarro_2018}). Here
we test the common and extreme
assumption that \textit{all SMGs at $z>4$ turn into a quiescent object
  at $3<z<4$} by matching their number densities
\citep[e.g.,][]{toft_2014}. For the calculation, we adopt a final
number density for quiescent galaxies at $3<z<4$ of
$n=1.4\times10^{-5}$ Mpc$^{-3}$ from S18b. We
compute results for the samples of SMGs with
optical/near-infrared counterparts from the ALESS
\citep{dacunha_2015}, S2CLS \citep{michalowski_2017}, and COSMOS
\citep{miettinen_2017} surveys. 

\subsubsection{Stellar masses and SFR distributions}
\label{subsubsec:distributions}
First, we can compare the distributions of the stellar masses and SFRs
of SMGs and the progenitors of quiescent galaxies. In Figure
\ref{fig:progenitors_properties} we present the (Gaussian) kernel density
estimation (KDE) of the observed distributions. We derived the expected
properties of the QG progenitors from the
SED modeling of SXDS-10017, SXDS-27434, and COS-466654 and the sample in
S18b. We adopted $\langle \mathrm{SFR} \rangle_{\rm
  main}$ as the average SFR during the main formation epoch and $50$\%
of the final stellar mass as $M_\star$ for the progenitors (excluding
mass loss and recycling, Section
\ref{sec:sed}). Notice that the choice of the SFH does not impact $\langle
\mathrm{SFR} \rangle_{\rm main}$ within the uncertainties (Figure \ref{figapp:sfhs}).
For the SMGs, we compiled the stellar masses from
\cite{dacunha_2015}, \cite{michalowski_2017}, and
\cite{miettinen_2017} and the SFRs they derived from the far-infrared
luminosities. Given the best-fit values for $\Delta t_{\rm form}$, the
far-infrared luminosities and $\langle \mathrm{SFR} \rangle_{\rm
  main}$ probe similar timescales for star formation. Not all the SMGs have an
optical/near-infrared counterpart, necessary to derive a stellar
mass. Therefore, the results are biased towards the less
dust-obscured objects, either at lower masses or more advanced stages
of formation, when the dust shrouding the cocoons of star formation
starts fading away.\\ 

By definition, the integral of each probability density curve is equal
to $\int_{-\infty}^{+\infty}P(x)\,dx=1$. In order to compare the various
populations, we normalized the KDE to the number
density of each sample ($\int_{-\infty}^{+\infty}n\,P(x) \,dx = n$
[Mpc$^{-3}$]). Therefore, the area under each curve in Figure
\ref{fig:progenitors_properties} is exactly equal to the number density of each sample,
with $\mathcal{C}_{\rm duty} = 1$ for the SMGs (no correction
for duty cycle).\\ 

The left sides of the distributions are in first approximation related
to the depth
of the observations. The $K_{\rm s}$ band limits
are similar for all the surveys and, indeed, the stellar mass
distribution for the quiescent galaxies and their progenitors, the
ALESS \citep{dacunha_2015} and the COSMOS surveys
\citep{miettinen_2017} show a consistent lower mass
limit. The only exception is the S2CLS survey \citep{michalowski_2017}, which -- even sharing a
similar $K_{\rm s}$ band depth with all the other surveys -- results
in a non-negligible probability of low stellar masses. However, this
tail of low-mass galaxies disappears when considering only $z_{\rm phot}$
estimates from the optical/near-infrared. Its existence may therefore be a spurious effect due
to wrong redshift estimates from the far-infrared SED, which we
therefore discarded. For clarity, we show the stellar mass distribution
without this low-mass tail in Figure \ref{fig:progenitors_properties}.
For what concerns the SFRs, it is evident the effect of the cut imposed by the sensitivity
limits of the SMG surveys, resulting in a lower limit on the $L_{\rm
  IR}\sim\mathrm{SFR}$. 
On the other hand, the lower limit of the
distribution of the progenitors mainly depends on the modeling of the
SED and the best SFH.\\ 

The peaks of the distributions of the stellar masses are 
consistent for the ALESS survey and the progenitors of the quiescent
galaxies, while the modeling of the S2CLS and the COSMOS sources
results in larger stellar masses than the quiescent (and
star-forming) population at lower redshift, as
already reported in \cite{miettinen_2017} and \cite{michalowski_2017}.
Under the initial assumption that all SMGs at $z>4$ become quiescent at
$3<z<4$, the massive SMGs cannot be considered only a tail of the
overall distribution (Appendix \ref{app:duty}). Notice that here we do not include the stellar
mass that the ongoing episode of star formation will add to SMGs,
which would further increase the discrepancy with the quiescent
population. On the other hand, all the distributions of SFRs roughly peak at the same
value, even if the KDE for the progenitors of quiescent galaxies
allows for significant probability densities at low SFRs, not being
hampered by an observational limit. The extension of the distribution
of the progenitors towards low SFRs points
towards \textit{the necessity of including less extreme systems 
to explain the existence of red, quiescent galaxies at
$3<z<4$}. Independent structural and dynamical arguments suggest a
similar conclusion on the origin of quiescent galaxies at
$z\sim2$, whose precursors might be compact blue SFGs with
properties typical of objects on the main sequence
\citep{barro_2013,
  barro_2017,popping_2017, gomez-guijarro_2019} or a dust-obscured
population so far overlooked \citep{williams_2019, wang_2019}. This shows the relevance of
pushing our search for DSFGs at lower SFRs, and not limit it to
classical starbursting SMGs. We come back to this point later, providing
supporting evidence from cosmological simulations (Section
\ref{sec:simulations}).

\subsubsection{The duty cycle of SMGs}
\label{sec:dutycycle}
As mentioned in Section \ref{subsubsec:smg_numberdensities}, only a
fraction of the global star-forming
population shines as SMGs at a specific time, this fraction
  depending on the sub-mm flux threshold to define an SMG. With the usual
assumption that all such systems at $z>4$ turn into the quiescent population at
$3<z<4$, we can therefore derive the duty cycle correction
$\mathcal{C}_{\rm duty}$ necessary to make the number densities of
the two populations exactly equal \citep[e.g.,][]{toft_2014}. Here we do this by
matching the number densities (i.e., the area under the curves in
Figure \ref{fig:progenitors_properties}) and the
stellar mass or SFR distributions (i.e., the shape of the
  curves) of the quiescent progenitors and the samples of
SMGs (see Appendix \ref{app:duty} for the details).
We derive $t_{\rm burst} \sim
200-400$~Myr for the sources in the ALESS survey. For the COSMOS and
S2CLS surveys we find similar $t_{\rm burst}\sim100$~Myr matching
the SFR distributions, but substantially shorter values down to
$t_{\rm burst}\sim10$~Myr when matching the \mstar\ KDEs. The values
for $t_{\rm burst}$ are consistent with
the typical depletion timescales of strongly star forming galaxies at
high-redshift ($t_{\rm duty}\sim100-200$~Myr, \citealt{daddi_2010,
  tacconi_2010, casey_2014}) and with the formation timescales via the
archeological approach \citep[e.g.,][]{thomas_2005, renzini_2006,
  choi_2014, onodera_2015}. However, at least for the self-consistent values of
the ALESS sources, we find duty cycle corrections
$\mathcal{C}_{\rm duty} \sim 5\times$ smaller than
estimated by \cite{toft_2014} for the range $2<z<3$. Notice that this
is partly due to the longer $t_{\rm burst}$ we estimate, and partly it
is a natural consequence of observing higher redshift sources as
the time spanned between $4<z<6$ is half of the interval between
$2<z<3$.

These calculations depend on a set of assumptions that
we specified at each step and they are affected by several sources of
uncertainties. An extended discussion is reported in Appendix
\ref{app:duty}. Here we stress once again that the evolutionary connection we tested here
relies on the extreme assumption that all SMGs at $z>4$ become
quiescent at $3<z<4$, driving to the introduction of a duty cycle
correction. Looser conditions (e.g., only $X$\% of the SMGs
turn into quiescent galaxies) result to the first order in longer duty cycles by a
similar amount. Similar rescaling factors apply when considering a
value different than our reference estimate of the comoving number
density of quiescent galaxies.

\section{Can cosmological simulations capture the formation of quiescent
galaxies at $\lowercase{z}\sim4$?}
\label{sec:simulations}

\subsection{Realistic comoving number densities of quiescent galaxies}
\label{sec:simulations_quenched}
Previous attempts of
reproducing the population of quenched galaxies at $z>3$ with cosmological
simulations and semi-analytical models fall short in producing enough systems by up to an order
of magnitude in the majority of cases \citep[S18b,][]{cecchi_2019}.
Here we explore the content of the recent \textit{Illustris} TNG
cosmological simulation public release \citep{marinacci_2018,
  naiman_2018, volker_2018, pillepich_2018, nelson_2018, nelson_2019},
in comparison with the previous 
\textit{Illustris} project (\citealt{vogelsberger_2014,
  vogelsberger_nature_2014, genel_2014, nelson_2015}; see
\citealt{merlin_2019} for a similar attempt).
As for the observations presented above, we
structured our search in two steps: we
first looked for the quiescent galaxies in the largest boxes available
TNG 300 ($205/h$ comoving Mpc) and TNG 100 ($75/h$ comoving Mpc), so
to build enough statistics of these rare
systems. We subsequently studied the
progenitors of the quenched systems possibly shining as SMGs in these boxes
\citep[C. Hayward et al. in preparation,][]{hayward_2013}. \textit{Illustris}
TNG 100 has a similar box size to the old \textit{Illustris}-1, so we
drew comparisons among these two to test the performances of the new simulations.\\  

\begin{deluxetable*}{lcccc}
  \tabletypesize{\normalsize}
  \tablecolumns{5}
  \tablecaption{Number and number densities of quiescent galaxies and their progenitors
    in the \textit{Illustris} and \textit{Illustris} TNG cosmological simulations.\label{tab:illustris}}
  \smallskip
  \tablehead{
    \colhead{}&
    \colhead{\textit{Illustris}-1}&
    \colhead{\textit{Illustris} TNG 100-1}&
    \colhead{\textit{Illustris} TNG 100-2}&
    \colhead{\textit{Illustris} TNG 300-1}\\
    \colhead{Box size [cMpc]}&
    \colhead{75/$h$}&
    \colhead{75/$h$}&
    \colhead{75/$h$}&
    \colhead{ 205/$h$}\\
    \colhead{Dark matter mass resolution [$10^6\,M_{\odot}$]}&
    \colhead{$6.3$}&
    \colhead{$7.5$}&
    \colhead{$59.7$}&
    \colhead{$59$}}
  \startdata
  \multicolumn{5}{c}{Quiescent galaxies\tablenotemark{a}}\\
      \rule{0pt}{3ex}  
      N$_{<-11}$, N$_{[-11,-10]}$ $(z=3)$&  1, 3& 45, 10&  25, 6& 631, 123\\
      $n_{<-11}$, $n_{[-11,-10]}$ $(z=3)$ [$10^{-6}$ Mpc$^{-3}$]&  0.8, 2.4& 36.6, 8.1$^\dagger$& 20.3, 4.9& 25.1, 4.9 \\
      \rule{0pt}{3ex}  
      N$_{<-11}$, N$_{[-11,-10]}$ $(z=3.7)$&  0, 1& 8, 0& 0, 1& 53, 12\\
      $n_{<-11}$, $n_{[-11,-10]}$ $(z=3.7)$ [$10^{-6}$ Mpc$^{-3}$]&  --, 0.8& 6.5$^\dagger$, --& --, 0.8& 2.1, 0.5\\
      \hline
      \multicolumn{5}{c}{Sub-millimeter galaxies\tablenotemark{b}}\\
      \rule{0pt}{3ex} 
      N$_{>\rm 1\,mJy}$, N$_{>\rm 3.5\,mJy}$ $(z=4)$ &  49, 0 & 51, 2 & 59, 2& 1224, 48 \\
      $n_{>\rm 1\,mJy}$, $n_{>\rm 3.5\,mJy}$ $(z=4)$ [$10^{-6}$ Mpc$^{-3}$]&  39.8, -- & 41.5, 1.6& 48.0, 1.6& 48.7, 1.9 \\
      \rule{0pt}{3ex}  
      N$_{>\rm 1\,mJy}$, N$_{>\rm 3.5\,mJy}$ $(z=5)$ & 11,  1& 15, 0 & 17, 1& 393, 11 \\
      $n_{>\rm 1\,mJy}$, $n_{>\rm 3.5\,mJy}$ $(z=5)$ [$10^{-6}$ Mpc$^{-3}$]& 8.9, 0.8& 12.2, --& 13.8, 0.8& 15.6, 0.4\\
      \enddata
      \tablenotetext{a}{Number of galaxies in the $z=3$ and $z=3.7$ snapshots with
        $M_\star \geq 4\times10^{10}$ \msun\ and
        $\mathrm{sSFR} \leq 10^{-11}$ yr$^{-1}$ (N$_{<-11}$, ``quenched'') or $10^{-11} <
        \mathrm{sSFR} \leq 10^{-10}$ yr$^{-1}$ (N$_{[-11,-10]}$,
        ``quenching''). All the quantities are computed within
        twice the stellar half-mass radius (\textsc{INRAD} in the
        \textit{Illustris} data
        releases). The comoving number densities $n$ of ``quenched'' and
        ``quenching'' galaxies, obtained dividing N$_{<-11}$,
        N$_{[-11,-10]}$ by the comoving volume of the box (Size$^3$,
        adopting $h=0.7$).}
      \tablenotetext{b}{Number of galaxies in the $z=4$ and $z=5$ snapshots with $S_{850}$ flux
     $>1$ mJy and $>3.5$ mJy within 25 kpc, computed following \cite{hayward_2013}, Hayward et al. (in
     preparation).\\
     $\dagger$ These values are not corrected by
       the factor $1.2\times$ for the volume of the box (Section \ref{sec:mass_resolution_effect}).} 
\end{deluxetable*}

To mimic our observational selection, we identified quiescent galaxies in the
$z=3.7$ snapshot based on the sSFR within twice the stellar half-mass
radius (\textsc{INRAD} quantities in the catalogs). The SFR is averaged over 10 Myr, but we checked for the consistent
quiescence in the descendant subhalos down to $z=2$, in order to
exclude contamination of temporary low-activity galaxies. We selected
both ``quenched'' and ``quenching'' (or post-starburst) galaxies fixing a 
threshold of $\mathrm{sSFR} \leq 10^{-11}$ yr$^{-1}$ and $10^{-11} <
\mathrm{sSFR} \leq 10^{-10}$ yr$^{-1}$, respectively, and imposing a minimum
$M_\star = 4\times10^{10}$ \msun\ similar to the mass completeness
limits in the observations (Section \ref{sec:number_densities_qg}).
Note that such a selection in sSFR is robust against variations of the
timescale over which the SF is averaged in simulations in the range $\sim10-200$
Myr and the measurement of quantities in different apertures
\citep{sparre_2015, davidzon_2018}. Moreover, the
  quiescent fractions are similar when separating galaxies based on
  colors or distance from the main sequence \citep{donnari_2019}.\\

We summarize the results of the search in Table \ref{tab:illustris}
and Figure \ref{fig:number_densities}. The comoving number densities $n$
of quenched galaxies in the \textit{Illustris} TNG boxes are
consistent only with the lowest observational estimates at $z=3.7$, while the old
\textit{Illustris} run does not contain enough of these objects, as
previously noted \citep[S18b]{wellons_2015} and in line with similar
previous attempts in smaller boxes \citep[e.g.,
\textsc{Mufasa},][]{dave_2016}. On the contrary, we retrieve a
numerous enough population of quiescent galaxies in the $z=3$ snapshot
(i.e., the lower limit of the redshift range explored here with spectroscopy) of the \textit{Illustris} TNG simulations, while the old \textit{Illustris} still fails
at reproducing the observed number densities. This is likely due to
the new feedback scheme implemented in the TNG simulations.

\subsection{The impact of the mass resolution}
\label{sec:mass_resolution_effect}
 Interestingly, the largest TNG 300 box
contains $2.5\times$ and $1.5\times$ less quenched/quenching galaxies
per unit comoving volume than TNG 100 at $z=3.7$ and $z=3$,
respectively (Table \ref{tab:illustris}). We checked for the effect of
the $\sim8\times$ lower mass resolution in TNG 300 than in TNG
100, which might not result in full convergence. As discussed in
\cite{pillepich_2018}, the lower resolution of TNG 300(-1) translates into lower
stellar masses and SFR than in TNG 100(-1), which might bias our number densities.
Therefore, we compared the number of galaxies above our mass threshold ($M_\star
\geq 4\times10^{10}$ \msun) in TNG 100-1 and TNG 100-2, the latter having a
resolution similar to TNG 300-1. 
We retrieve $\sim25$\% less galaxies in TNG 100-2 than in TNG 100-1, this fraction
increasing when selecting quenched/quenching galaxies, so that we find
only 1 (31) objects with $\mathrm{sSFR}\leq10^{-10}$ yr$^{-1}$ at
$z=3.7$ ($z=3$) in TNG 100-2, a factor of $8\times$ ($1.7\times$)
less than TNG 100-1. Therefore, while at $z=3.7$ the low number
statistics and the cosmic variance likely dominate the difference between
TNG 300-1 and TNG 100-1, the mass resolution explains the
discrepancy at $z=3$ for our selection.\\

\noindent
From the comparison between the number densities
  of quenched/quenching objects in TNG 300-1 and TNG 100-2, we derive
  a simple factor to correct the comoving number densities from the
  smaller box of the TNG 100-1 run, while retaining the advantages of
  the high resolution. We compute such correction as the ratio of the
  number densities $n_{\rm TNG300}/n_{\rm TNG100(2)} = 1.2\times$ at
  $z=3$, where we have enough statistics, and we assume that this factor applies also
  at the previous snapshot at $z=3.7$. Therefore, our fiducial number densities of
  quenched/quenching objects for the TNG 100-1 box are
  $5.4\times10^{-5}$ and $7.8\times10^{-6}$ Mpc$^{-3}$ at $z=3-3.7$,
  respectively.

\subsection{Simulated sub-mm galaxies and their properties}
\begin{figure}
  \centering
  \includegraphics[width=\columnwidth]{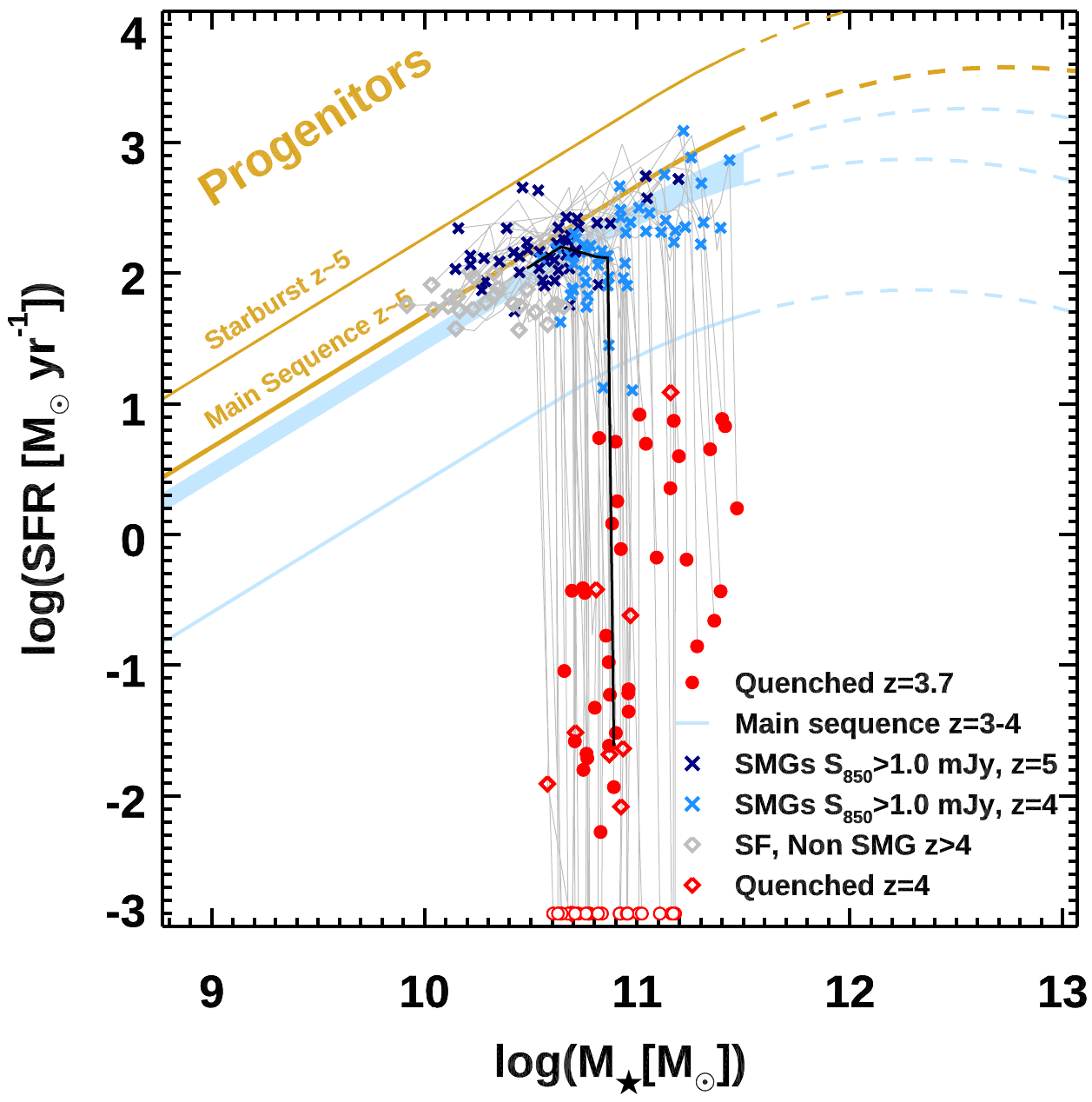}
  \caption{\textbf{Stellar mass and SFR plane for the
      \textit{Illustris} TNG-300 simulation.} The red
  circles indicate the location of quiescent galaxies
  ($\mathrm{sSFR}\leq10^{-10}$ yr$^{-1}$ within twice the half-mass
  radius) in the $z=3.7$ snapshot (the  empty symbols represent upper
  limits fixed at $\mathrm{SFR}=10^{-3}$
  \myr\ for display purpose). The location
  of the main sequence at $z=3-4$ as parametrized in
  \citet{schreiber_2015} is shown by the blue shaded
  area. The blue solid line marks the position of sources $10\times$ below
  the main sequence at $z=3-4$. The golden solid lines indicate the
  position of the main sequence and $4\times$ above it at $z=5\,(
  \sim z_{\rm form})$. The dashed lines show the extrapolation of the main
  sequence to masses larger than $\mathrm{log}(M_\star/M_\odot) =
  11.5$ in order to facilitate the comparison with Figure \ref{fig:ms}.
 Dark and light blue crosses mark SMGs with $S_{850\rm \mu
    m}\geq1$ mJy at $z=5$ and $z=4$, respectively, as computed by Hayward
  et al. (in preparation). 
The gray and red diamonds show SFGs with $S_{850\rm \mu
    m}\leq1$ mJy at $z>4$ and quiescent objects at $z=4$, respectively.
 The gray tracks show the evolution of
  $z=3.7$ quiescent objects back to $z=5$, while the black solid line
  shows the median evolution in
  the same interval.}
  \label{fig:ms_illustrisTNG}
\end{figure}
 We then traced the evolution of the main subhalo progenitor of each
quiescent galaxy back to the main formation epoch derived for
the observed targets ($z\sim5$, Section \ref{sec:sed}). In Figure
\ref{fig:ms_illustrisTNG} we show the evolutionary tracks in the
\mstar-SFR plane for TNG-300, but similar conclusions hold for 
TNG-100. In order to directly compare with the operational
definition of SMGs widely spread among observers, we also show the
modeling of the 850 $\mu$m emission of $z=4-5$ galaxies by
Hayward et al. (in preparation). The authors compute mock 850 $\mu$m fluxes for
all the sources with $M_\star>4\times10^{9}$ \msun, based
on a relation derived from full radiative transfer calculations on
idealized discs and mergers at a spatial resolution inaccessible for a
cosmological simulations \citep{hayward_2011, hayward_2013}. The
calculation is performed by integrating the ISM and the SFR within 25 kpc
(and not within twice the half-mass radius) excluding the particles in satellites, in order
to compare with typically resolved SMGs (Hayward et al. in prep.).
A significant difference between SFR(25 kpc)
and SFR(INRAD) is the reason why a minority of objects display bright
$S_{850}$ fluxes while being formally quiescent in our selection
(Figure \ref{fig:ms_illustrisTNG}). 
Notice that the 850~$\mu$m emission depends not only on the SFR, but also on the dust mass and
its temperature: highly star-forming galaxies might thus not shine as
SMGs due to a lower dust content and, thus, hotter dust temperature
\citep[e.g.,][]{hayward_2012,safarzadeh_2016}. As a reference, we adopt here cuts at
$S_{\rm 850 \mu m}>1$ mJy and $>3.5$ mJy, comparable with the ALESS and
the COSMOS/S2CLS limiting sensitivities, respectively.\\

While differing in principle, the SMGs have similar \mstar\ and SFR to
the progenitors of the quenched galaxies at $z=3-3.7$. However, they
are consistent with being the tail of most massive galaxies on the
main sequence at their redshift, without reaching the extreme values of observed
starbursts (Figure \ref{fig:ms}). The choice of the $S_{850}$ flux
thresholds has a strong impact on the stellar mass of the selected
SMGs, while the redshift does not. Flux cuts at $>1$ and $>3.5$ mJy
results in median $\mathrm{log}(M_\star/M_\odot) = 10.5$ and $11.1$,
respectively, both at $z=4$ and $z=5$ in TNG 300. This naturally
arises from the fact that, by definition, a higher $S_{850}$ threshold
selects for higher SFR and $M_{\rm dust}$ values, both of which
correlate with $M_\star$. Therefore, a $S_{\rm 850}$-\mstar\ relation
is in place at least for main-sequence galaxies, matching the
observations \citep{hayward_2013}.

\subsection{Number densities of simulated sub-mm galaxies}
We further calculated the number densities for the SMGs with $S_{\rm
  850}>1$ and $>3.5$ mJy in the snapshots
at $z=4$ and $z=5$ (Table \ref{tab:illustris} and Figure
\ref{fig:number_densities}).  To avoid
overpopulating the latter, we show only the number densities for a
threshold of $S_{\rm 850} = 1$ mJy, similar to the ALESS limiting
flux. At face value, the simulations slightly
overshoot the empirical number densities uncorrected for the duty
cycle, but we find an overall good agreement within the
observational uncertainties. Notice that no correction is needed for
the simulated number densities.
Similar conclusions hold when comparing
simulations with the COSMOS/S2CLS surveys at
the corresponding depth. The comoving number
densities for SMGs with $>1$ mJy are all consistent within 20\% among the three
suites of simulations, while we find a sizable sample of objects with
$>3.5$ mJy only in TNG-300. We further checked for the dependence on the
mass resolutions of the simulations as for the quenched galaxies
(Section \ref{sec:mass_resolution_effect}) by comparing TNG 100-1 and
TNG 100-2, which has a similar resolution to TNG 300-1. We retrieve
$\sim10-15$\% less galaxies with $S_{850}$ fluxes $>0.1-0.5$ mJy, but
$\sim15-30$\% more SMGs in TNG 100-2 at $z=3-5$ than in TNG 100-1. 
The low number statistics dominate the comparison for higher $S_{850}$
flux thresholds. These fractions should be considered as the typical
uncertainties on the number densities we derived for TNG
300-1. We also attempted a calculation of a volume
  correction for
  the SMGs in TNG 100-1 by comparing TNG 300 and TNG 100-2 (Section
  \ref{sec:mass_resolution_effect})
  at $z=4-5$ down to our reference flux threshold of
  $S_{\rm 850} = 1$ mJy. However, such factor is negligible
  ($2-10$\%).\\

Moreover, Hayward et al. (in
preparation) find that \textit{Illustris} TNG systematically
underpredicts the $850$ $\mu$m cumulative number counts (therefore,
integrated over redshift) with respect to
observations and the old \textit{Illustris} run at similar mass resolution, owing to the
increased efficiency of AGN feedback reducing the number of massive
strongly star-forming systems. The difference between the two
simulation sets is not evident at $z\geq4$. This is likely due to the
fact that SMGs at these redshifts have not reached yet the critical
black hole mass to trigger efficient AGN feedback. Therefore, the
discrepancy between \textit{Illustris} and \textit{Illustris} TNG
reported by Hayward et al. is due to the later evolution, while
systems at $z>4$ constitute only a
minor fraction of the SMG population.

\subsection{Not all progenitors are sub-mm galaxies}
The comparison between the comoving number densities of quiescent
galaxies at $z=3-3.7$ and the SMGs at $z=4-5$ (Table
\ref{tab:illustris}) shows that \textit{only a fraction of the
  SMGs at high redshift turn quiescent at $z=3-3.7$, this fraction
  depending on the sub-mm flux threshold.} In TNG-300,
$88-90$\% of the SMGs with $S_{850}>3.5$ mJy at $z=4-5$ are quenched
($\mathrm{sSFR}<10^{-10}$ yr$^{-1}$) at $z=3$, but only $20-30$\% at
$z=3.7$. These fractions drop significantly when considering a lower
flux threshold of $S_{850}>1$ mJy: $45-60$\% of the SMGs above this
limit at $z=4-5$ are quenched at $z=3$ ($5-15$\% at $z=3.7$). We do
not attempt the comparison in TNG-100 and \textit{Illustris}-1 due to
the low or even absent statistics. The
difference due to the flux threshold likely mirrors the 
mass selection mentioned above: larger $S_{850}$ fluxes correspond to
larger masses, therefore closer to the threshold to ignite efficient
AGN feedback and a rapid quenching. The shorter time interval between 
$z=4$ and $z=3.7$ concurs to the drop of the fractions of SMGs
quenching between these two limits, compared with quenching occurring
in the $z=3-5$ period. Finally, at the spatial and time resolution
\textit{Illustris} TNG, lower $S_{850}$ appear to be sustainable for longer timescales. 
Dedicated simulations at higher resolution, necessary to capture the
stochasticity of the processes igniting the brightest SMGs, could test
this result, but they are beyond the scope of this paper.\\ 

Inverting the order of the terms of comparison, we find that
\textit{only a fraction of the progenitors of quenched galaxies at
  $z=3-3.7$ shine as SMGs at $z=4-5$, this fraction depending on the
  sub-mm flux threshold.} As Figure \ref{fig:ms_illustrisTNG} shows,
$80$\% of the quenched galaxies at $z=3.7$ have 
$S_{850} > 1$ mJy at $z=4$ ($70$\% at $z=5$) in TNG-300, while $12$\%
($0$\%) were already quenched at $z=4$ ($z=5$).
The fraction of sub-mm bright progenitors drops when considering higher
flux thresholds: $14$\% of the
quiescent galaxies at $z=3.7$ have $S_{850} > 3.5$ mJy at $z=4$ ($5$\%
at $z=5$).\\

This comparison suggests a more complex connection between the
progenitors of high-redshift quiescent galaxies and SMGs than previously
assumed (e.g., Section \ref{sec:evolutionary}). While the majority of extremely bright SMGs at $z=4-5$ detectable by
shallow surveys quench by $z=3$, a substantial fraction of the highest
redshift quenched systems have less extreme progenitors, lying on the
main sequence during their epoch of main stellar mass assembly and
emitting sub-mm fluxes partially detectable only by the deepest
surveys currently available.

\section{Conclusions}
\label{sec:conclusions}

In this work we reported the discovery and the detailed analysis of
three quiescent galaxies at 1.5 Gyr after the Big Bang. Dedicated
spectroscopic follow-up in the optical/near-infrared allowed us to
securely confirm the redshift of two of the sources ($z=3.775, 4.012$) and to put
tentative constraints on the third one ($z\approx3.767$). Their quiescence is supported by
the modeling of their SED, by the absence of emission lines, and the 
non-detections in the far-infrared/sub-millimeter regimes. Given their
large stellar masses of $\sim10^{11}$ \msun, these objects are
located $\gtrsim 1-2$ dex below the main sequence at their redshifts.
The combined modeling of the SED and the spectra suggests that these
galaxies went through a short phase ($\sim50$ Myr) of intense star
formation ($\sim1200-3500$ \myr) peaking $\sim150-500$ Myr prior the
time of observation, followed by an abrupt decrease of their SFRs and
the cessation of any relevant formation of new stars. 

We then explored their connection with star-forming progenitors at
higher redshifts, testing previous suggestions of a direct link with
strongly starbursting SMGs. We compared the comoving number densities
$n$ of 850~$\mu$m and/or 1.1 mm-selected
SMGs at $z>4$ and quiescent galaxies at $3<z<4$, compiling recent
results in the literature. In general, a large scatter affects the
estimates of $n$ for both populations, predominantly due
to a combination of factors dominated by different classification
schemes and selection criteria and by the uncertainties on the
redshift. We find the number densities of SMGs at $z>4$ from the
deepest surveys ($S_{\rm 850\,\mu m}\gtrsim1$ mJy) to be in broad
agreement with the estimates for quiescent galaxies at
$3<z<4$. Brighter SMGs ($S_{\rm 850\,\mu m}\gtrsim3.5$ mJy) are $6-20\times$
less numerous. Adopting the assumption of a univocal
  correspondence between SMGs and the progenitors of quiescent
  galaxies, we attempted to estimate the duty cycle correction necessary
to match their comoving number densities. The resulting duty cycle is of the
order of the depletion timescale estimated for strongly star-forming
galaxies at high-redshift ($t_{\rm burst}\sim100-400$ Myr), but
this value significantly drop to $\sim 10$ Myr for the brightest SMGs,
questioning the underlying assumption. This is reinforced by the
comparison of the stellar masses and SFRs of SMGs and quiescent galaxies. We
find that the current SED modeling tends to overpredict \mstar\ for the
brightest SMGs with respect to quiescent galaxies at lower redshift,
while objects with lower sub-mm fluxes can better reproduce the
distributions of \mstar\ and SFR expected for the progenitors of quiescent
galaxies. This points towards the necessity of including less
  extreme systems to explain the existence of red, quiescent galaxies
  at $3 < z < 4$.

Finally, we tested our assumptions on the evolutionary path of
high-redshift quiescent galaxies by comparing our results with the
recent \textit{Illustris} TNG simulation. We retrieve comoving number densities of massive quiescent
galaxies at $z=3$ that are in fair agreement with the broad range of 
observed values, surpassing the performance of the previous generation
of cosmological simulations. This is especially due to the large box size
necessary to study these rare systems and the improved
  feedback scheme. However, we report a growing
inconsistency in the comoving numbers densities at increasing
redshift, so that at $z=3.7$ the simulated populations match only
the lowest among the observational estimates. Moreover, we traced their progenitors at $z=4-5$ and 
their mock sub-mm fluxes, finding a population of SMGs as numerous as in
the observations. We find that $\sim90$\% of the SMGs with
$S_{850}>3.5$ mJy at $z=4-5$ are quenched by $z=3$, but this fraction
drastically decreases to $\sim45-60$\% for dimmer sources with $S_{850}>1$ mJy.
In other words, we showed that not all $z=3-3.7$ quiescent
galaxies have an SMG progenitor and, similarly, that not all
SMGs at $z=4-5$ quench by $z\sim3-3.7$, the fractions mainly depending
on the sub-mm flux cut to select the dusty star-forming galaxies.
Moreover, as suggested by the observations mentioned above, simulations indicate that the
progenitors of massive quiescent galaxies at $z>3$ are not necessarily
prototypical extreme starbursting systems, but more
normal star-forming galaxies at the most massive end of the main sequence. This highlights the
importance of obtaining deep sub-mm observations of
high-redshift galaxies, abandoning the original definition of ``SMGs''
to focus on more physically meaningful categories.

\section*{Acknowledgements}
We acknowledge the constructive comments from the
  anonymous referee that improved the content and
  presentation of the results. We warmly thank Corentin
Schreiber for the useful discussions and guidance to the
use of \textsc{Slinefit} and \textsc{Fast++}. FV thanks Davide
Martizzi for discussions about the \textit{Illustris} simulations;
Tao Wang, and Helmut Dannerbauer for 
  comments about the sub-mm galaxy samples.
FV and GM acknowledge the Villum
Fonden research grant 13160 ``Gas to stars, stars to
dust: tracing star formation across cosmic time'', and the Carlsberg
Fonden research grant CF18-0388 ``Galaxies: Rise And Death''. ST, MS,
CGG, and GM acknowledge support from the European Research Council
(ERC) Consolidator Grant funding scheme (project ``ConTExt'', grant
number: 648179). The Cosmic Dawn Center (DAWN) is funded by the Danish
National Research Foundation under grant No. 140. 
MO acknowledges support from JSPS KAKENHI Grant Number JP17K14257.
MS acknowledges support by the European Research Council under ERC-CoG
grant CRAGSMAN-646955.
OI acknowledges the funding of the French Agence Nationale de la Recherche for the project ``SAGACE'' and the Centre National d'Etudes Spatiales (CNES).
KY was supported by JSPS KAKENHI Grant Number JP18K13578.
The Flatiron Institute is supported by the Simons Foundation.
Based on observations collected at the European Southern Observatory
under ESO programme 0100.B-0922(A).
Some of the data
presented herein were obtained at the W. M. Keck Observatory, which is
operated as a scientific partnership among the California Institute of
Technology, the University of California and the National Aeronautics
and Space Administration. The Observatory was made possible by the
generous financial support of the W. M. Keck Foundation. 
The observations were carried out within the framework of Subaru-Keck
time exchange program, where the travel expense was supported by the
Subaru Telescope, which is operated by the National Astronomical
Observatory of Japan.
The authors wish to recognize and acknowledge the very significant cultural role
and reverence that the summit of Maunakea has always had within the
indigenous Hawaiian community. We are most fortunate to have the
opportunity to conduct observations from this mountain.

\appendix
\section{Stellar population models, star formation histories, and
  their parametrization}
\label{sec:appendix}
\begin{figure*}
\centering
\includegraphics[width=\textwidth]{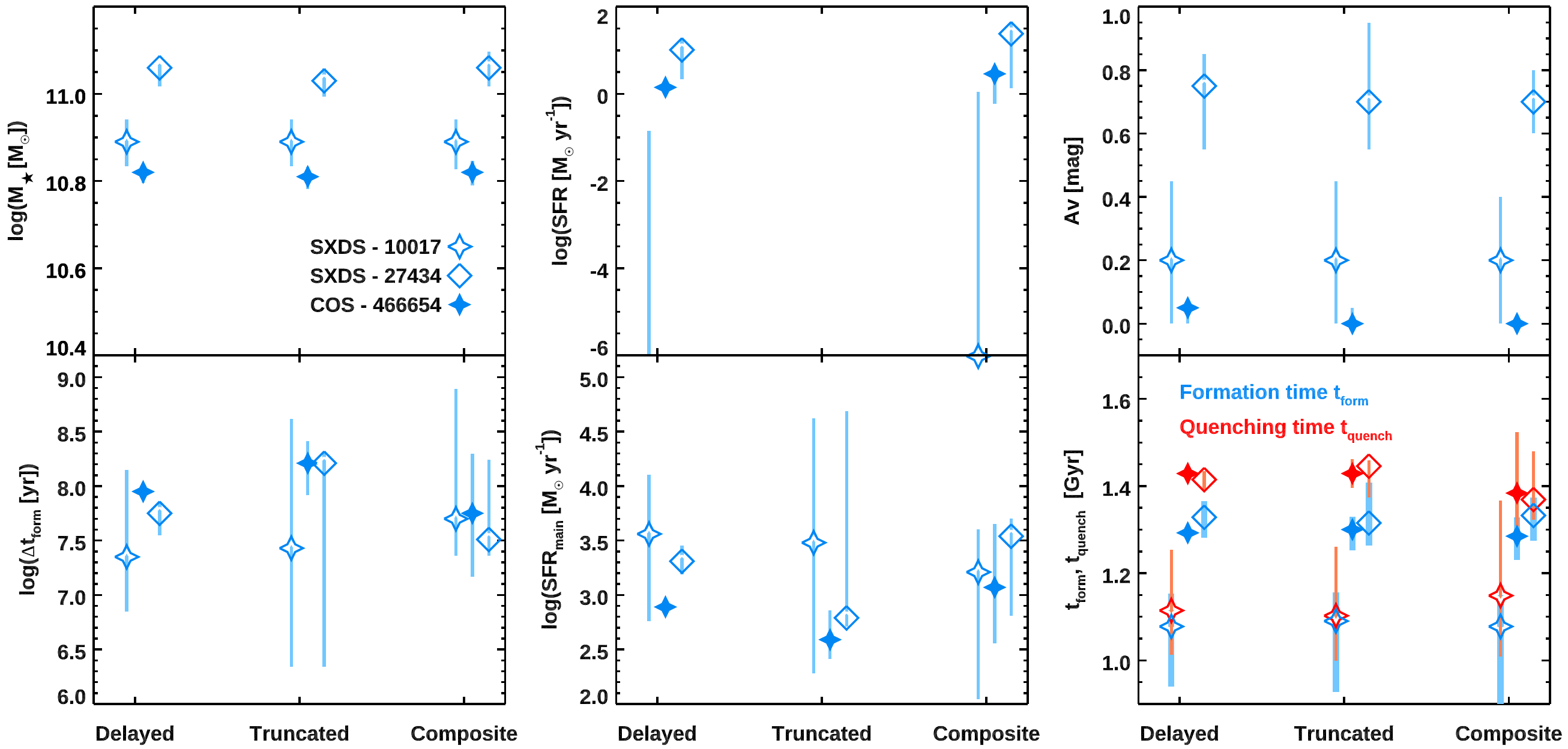}
\caption{\textbf{Main quantities derived from SED modeling varying the
  SFH.} From top left (clockwise): Stellar mass, SFR, $A_{\rm V}$ dust
attenuation, $t_{\rm quench}$, $t_{\rm form}$, $\langle \mathrm{SFR}
\rangle_{\rm main}$, and duration of the main formation epoch as a
function of the SFH (delayed, truncated, or composite as in
\citealt{schreiber_2018c}) for the three sources studied in this work (open
star: SXDS-10017; open diamond: SXDS-27434; filled star: COS-466654). The definition of each
quantity and the analytical expressions of the SFHs are described in
Section \ref{sec:sed}. The error bars corresponds to the 90\%
confidence intervals computed following \citet{avni_1976}.}
\label{figapp:sfhs}
\end{figure*}
\begin{figure*}
\centering
\includegraphics[width=\textwidth]{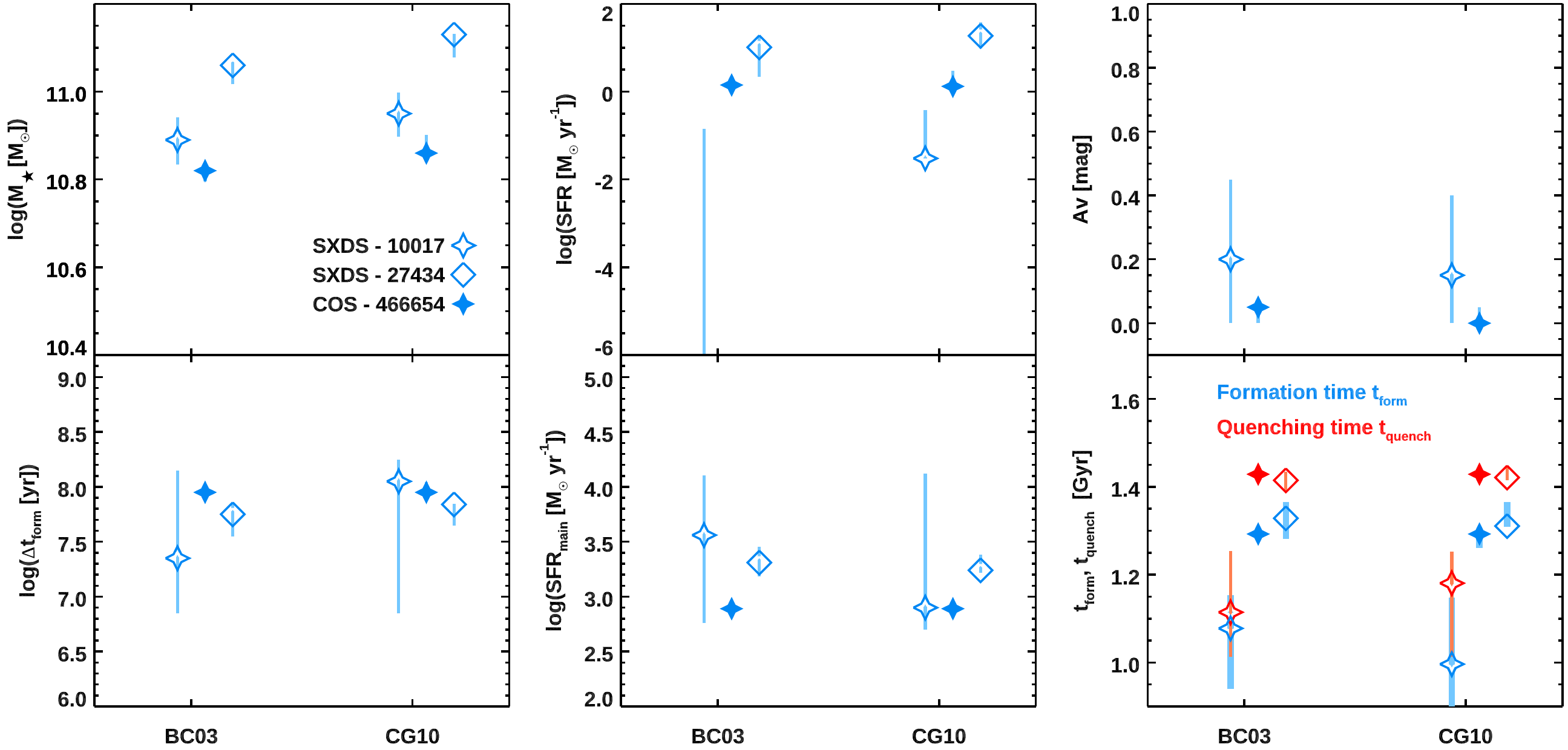}
\caption{\textbf{Main quantities derived from SED modeling varying the
  stellar models.} From top left (clockwise): Stellar mass, SFR, $A_{\rm V}$ dust
attenuation, $t_{\rm quench}$, $t_{\rm form}$, $\langle \mathrm{SFR}
\rangle_{\rm main}$, and duration of the main formation epoch as a
function of the stellar population models (Bruzual \& Charlot 2003,
Conroy \& Gunn 2010) for the three quiescent galaxies at fixed delayed
SFH. The error bars corresponds to the 90\%
confidence intervals computed following \citet{avni_1976}.}
\label{fig:stelpop_comparison}
\end{figure*}
In Figure \ref{figapp:sfhs} we show the main quantities derived from
the SED modeling that we used in this work, obtained by fitting the same
set of data, but varying the parametrization of the SFH. The quantities
we adopted to describe the formation and quenching epochs of the
studied objects are robust against the choice of the SFH. Notice that
by definition the current SFR for the truncated SFH is null. The error
bars represent the 90\% confidence intervals computed following the
$\chi^2$ criterion by \citealt{avni_1976} (Section
\ref{sec:sed}). In Figure
  \ref{fig:stelpop_comparison} we present the same quantities
  computed by varying the stellar templates \citep{bruzual_2003,
    conroy_2010}, while fixing the SFH to a delayed model. The choice
  of the stellar templates does not affect our results.

\section{Properties of candidate progenitors of quiescent galaxies}
\label{app:duty}

In Figure \ref{fig:progenitors_properties} we presented the (Gaussian) kernel density
estimation (KDE) of the observed stellar mass and SFR distributions of
quiescent galaxies at $3<z<4$ and SMGs at $z>4$, each one normalized by
  the comoving number density of the corresponding sample. In the central panels
of Figure \ref{fig:progenitors_properties_full}, we normalized
all the distributions to the same final number density
of quiescent galaxies, $n=1.4\times10^{-5}$ Mpc$^{-1}$ (S18b). This
value is only for reference, as the final results scale proportionally with
the choice of this parameter (Section \ref{sec:dutycycle}). To facilitate the
comparison and appreciate the difference between the different
normalizations, we show the same panels of Figure
\ref{fig:progenitors_properties} also in Figure \ref{fig:progenitors_properties_full}.
The shape of the KDEs is identical in the left and central panels, but the
areas are different.\\ 

First, the renormalization exacerbates the tensions among the
stellar mass distributions discussed in Section
\ref{subsubsec:distributions}, pointing at inconsistencies in the
modeling of the SMG population yet to be solved. Second, as mentioned in Section
\ref{sec:dutycycle}, we can derive a duty cycle for SMGs by assuming
that all SMGs at $z>4$ become quiescent at $3<z<4$.
The right panels of Figure \ref{fig:progenitors_properties_full} show the $t_{\rm burst}$ necessary
to match the number density (i.e., the area under the curves) and the
stellar mass or SFR distributions (i.e., the shape of the
  curves) of the quiescent progenitors and the samples of
SMGs. Matching the distributions of stellar masses or SFRs provide
consistent results for the ALESS survey as a consequence of the overall agreement of both set
of distributions with the progenitor properties. On the other hand, as noted above, 
matching the \mstar\ KDEs for the COSMOS and S2CLS sources results in
$>10\times$ shorter $t_{\rm burst}$ than matching
the SFR distributions. Notice that
the optical/near-infrared SED modeling to derive the stellar masses
for SMGs is in principle rather independent of the derivation of the
SFRs, the latter being connected to the far-infrared portion of the
spectrum, allowing us to separate the two approaches to estimate
$t_{\rm burst}$. 

\subsection{Caveats}
\begin{figure*}
\includegraphics[width=\textwidth]{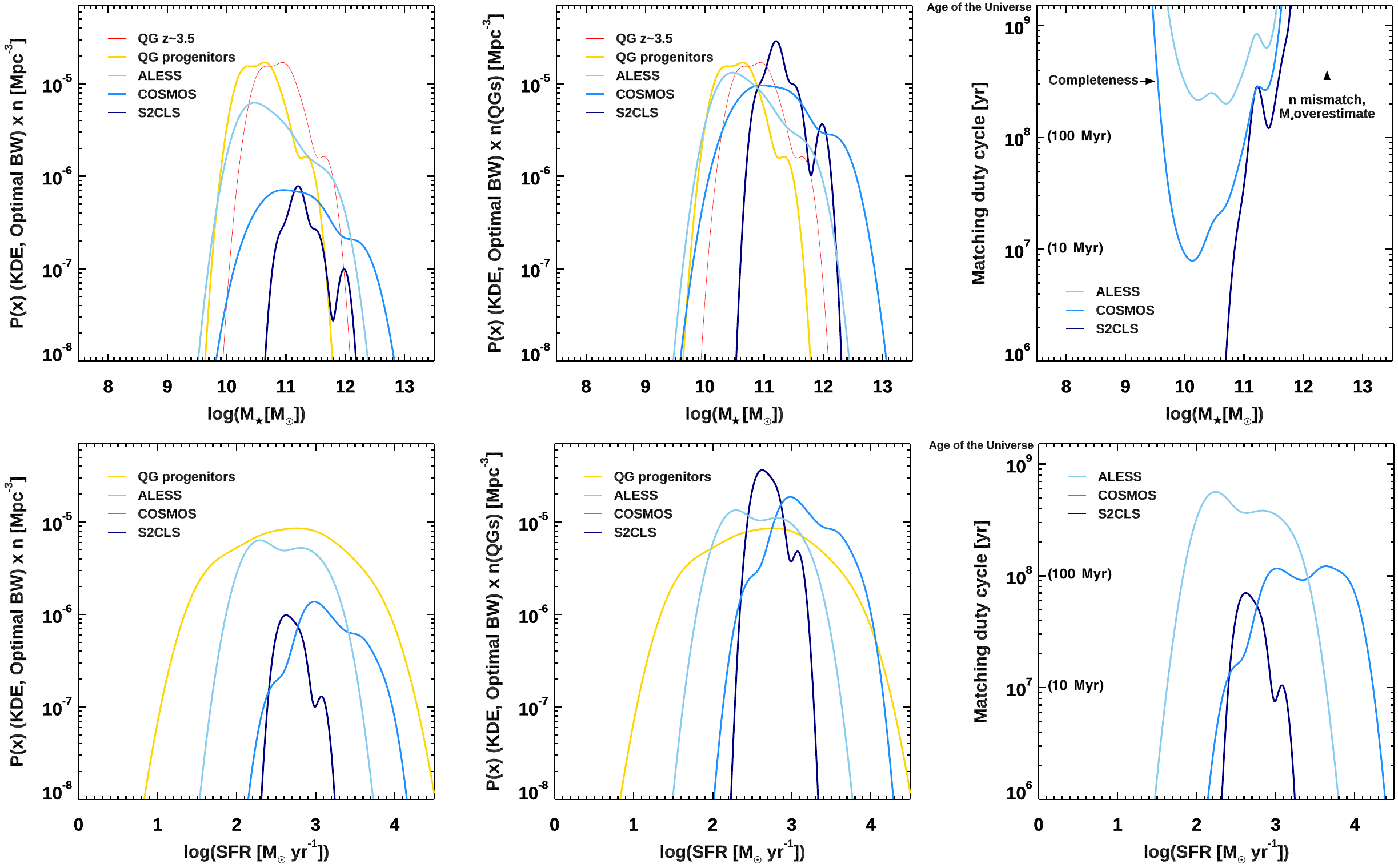}
\caption{\textbf{Stellar masses, star formation rates, and duty cycle of
    candidate progenitors of high-redshift quiescent galaxies.} \textit{Left:} Kernel Density
  Estimator (KDE) of the density probability distributions of stellar
  masses (\textit{top panel}) and SFR (\textit{bottom panel}) for 
  the progenitors at $z_{\rm form}$ of massive quiescent galaxies at $3<z<4$ (this
  work+S18b, golden line) and SMGs at
  $z>4$ from the ALESS, COSMOS, S2CLS surveys \citep[blue lines]{dacunha_2015, miettinen_2017,
    michalowski_2017}. For reference, we show the stellar mass
  distribution of the QGs (red thin line). The areas under the curves are normalized to
  the number density of each population (Figure \ref{fig:number_densities}). \textit{Center:} KDE of the
  distributions of stellar masses (\textit{top panel}) and SFR
  (\textit{bottom panel}) for the same populations of the left
  panel. The area under every curve is normalized to
  the number density of the quiescent population at $3<z<4$ by
  \citet{schreiber_2018c} (i.e., assuming that all SMGs at $z>4$ turn into
  quiescent galaxies at $3<z<4$). \textit{Right:} The blue color lines indicate
  the $t_{\rm burst}$ to assume as a function of \mstar\ to match the
  normalization (i.e., the number density) and the shape of
  the stellar mass (\textit{top panel}) and SFR (\textit{bottom
    panel}) distributions of high-redshift SMGs \citep{dacunha_2015, miettinen_2017,
    michalowski_2017} and the quiescent descendants at $3<z<4$ (this
  work+S18b), assuming that all SMGs become
  quiescent.}
\label{fig:progenitors_properties_full}
\end{figure*}

We mentioned that the derivation of the duty cycle correction is valid
only under the assumption that all SMGs at $z>4$ become
quiescent at $3<z<4$, but more hypotheses and limitations enter this analysis.
First, the calculation of the number densities depend
on the depth and completeness
of the surveys, and on the redshift estimates. In particular, the
latter are hard to obtain for SMGs, relying on the correct identification
of a optical/near-infrared counterpart and being naturally complicated
by the extreme obscuration of these sources. However, our attempt to
include the photometric redshift information and its uncertainty
represents an improvement towards the determination of reliable number
densities, rather than arbitrarily rescaling luminosity functions and
the ensuing number counts by fixed fudge factors (see Section \ref{sec:simulations}). The comparison
of the \mstar\ distributions of SMGs
and quiescent galaxies suffers from similar uncertainties, relying on
the modeling of the optical/near-infrared part of the
spectrum. Moreover, the details of the modeling and the assumptions behind
it notoriously generate systematic differences among different
results. An accurate reconstruction of the stellar masses of SMGs is
beyond the scope of this paper and we, thus, assumed results present in the
literature (see \citealt{gomez-guijarro_2018} for a detailed assessment of
the optical/near-infrared properties of some of the $z\sim4.5$ SMGs in
\citealt{miettinen_2017} and about the importance of spatial
high-resolution observations). Here we stress only two salient differences among the
various samples we presented. First, the use of different stellar
population synthesis codes for the quiescent population and the
SMGs. In particular, the SEDs of all the SMGs have been all modeled
with \textsc{Magphys}, which enforces a global energy balance between
the UV/optical and the infrared part of the electromagnetic
spectrum. On the other hand, the quiescent galaxies, including the three
sources presented in this paper, have been modeled with classical
codes not assuming any energy balance (\textsc{Fast++, LePhare, Mizuki}), but this should not play a major role for dust-poor quiescent
galaxies. The choice of the SFH is a
second element of relevance when using \textsc{Magphys} for SMGs,
driving variations up to $\sim0.5$ dex in the final stellar mass, as
evident from Figure \ref{fig:ms} and \ref{fig:progenitors_properties_full}
\citep{michalowski_2014}. Overall we find that the current
estimates of \mstar\ for SMGs are systematically larger than in lower
redshift galaxies, quiescent or not (see also \citealt{miettinen_2017} and
\citealt{michalowski_2017}). Notice that we did not take into account the contribution
of the stellar mass assembled during the ongoing episode of star formation
in SMGs at $z>4$, which would even increase the differences with the lower
redshift objects. We further remark that also the expected
properties of the progenitors of quiescent galaxies fully depend on
our implementation of the SED modeling, being therefore susceptible to
possible systematics due to the current assumptions. Finally, our calculations
assume fixed comoving number densities at the explored
redshifts, while mergers might a priori change the mass ranking at the
base of this technique \citep{torrey_2017}. 

\bibliography{bib_passive_valentino.bib} 

\end{document}